\documentclass[12pt]{article}

\usepackage{amsmath,amsthm,amssymb, amsfonts} 
\usepackage[dvipsnames]{xcolor}
\usepackage[top=1.1in, bottom=1.1in, left=1in, right=1in]{geometry}
\usepackage[onehalfspacing]{setspace}
\usepackage{natbib}
\usepackage{graphicx}
\usepackage{caption}
\usepackage{subcaption} 

\usepackage{algorithm}
\usepackage{algpseudocode}

\usepackage[colorlinks=true,linkcolor=RawSienna,citecolor=RawSienna,urlcolor=RawSienna,bookmarksopen=true,pdfstartview=FitB]{hyperref}

\newtheorem{theorem}{Theorem}[section]

\newtheorem{lemma}{Lemma}[section]
\newtheorem{proposition}{Proposition}[section]

\begin{document}
 
 
\newcommand{\blind}{0} 

\title{Weak Signal Inclusion Under Sparsity and Dependence}
\author{X. Jessie Jeng and Yifei Hu \\
	\vspace{0.1in}
	Department of Statistics, North Carolina State University}

\if0\blind
{
	\title{\bf Weak Signal Inclusion Under Sparsity and Dependence \thanks{Address for correspondence: X. Jessie Jeng, Department of Statistics, North Carolina State University, SAS Hall, 2311 Stinson Dr., Raleigh, NC 27695-8203, USA. E-mail: xjjeng@ncsu.edu. 
		}\hspace{.2cm}}
	\author{X. Jessie Jeng and Yifei Hu \\ 
		Department of Statistics, North Carolina State University\\
	}
	\maketitle
} \fi

\if1\blind
{
	\bigskip
	\bigskip
	\bigskip
	\begin{center}
		{\LARGE\bf Weak Signal Inference With False Negative Control Under Dependence and Sparsity \\}
	\end{center}
	\medskip
} \fi

\begin{abstract}
We consider the scenario where important signals are not strong enough to be separable from a large amount of noise. Such weak signals commonly exist in large-scale data analysis and play vital roles in many biomedical applications. Existing methods however are mostly underpowered for such weak signals. We address the challenge from the perspective of false negative control and develop a new method to efficiently regulate the false negative proportion at a user-specified level. The new method is developed in a realistic setting with arbitrary covariance dependence between variables. We calibrate the overall dependence through a parameter whose scale is compatible with the existing phase diagram in high-dimensional sparse inference. Utilizing the new calibration, we asymptotically explicate the joint effect of covariance dependence, signal sparsity, and signal intensity on the proposed method. We interpret the results using a new phase diagram, which shows that the proposed method can efficiently retain a high proportion of signals even when they cannot be well-separated from noise. Finite sample performance of the proposed method is compared to those of several existing methods in simulation studies. The proposed method outperforms the others in adapting to a user-specified false negative control level. We apply the new method to analyze an fMRI dataset to locate voxels that are functionally relevant to saccadic eye movements. The new method exhibits a nice balance in identifying functional relevant regions and avoiding excessive noise voxels.
\medskip
	

\end{abstract}

\section{Introduction}   

False negative errors are of major concern in many scientific inquires. 
In disease surveillance, failures to detect an outbreak at early stage may cause delayed action and a widespread occurrence.  
For neurosurgical patients, mistakenly deeming a brain region to be functionally uninvolved and subsequently resecting tissues that are vital to quality of life can cause significant harm. 
In human genetics, genome-wide association studies (GWASs), facilitated by high-throughput sequencing technologies, have revolutionized the field.  
Nevertheless, the identified GWAS variants for many complex traits have explained only a small percentage of trait hertiability (the ``missing heritability" problem), and as sample sizes for GWAS continue to grow, the effect sizes of newly identified variants are mostly very weak \citep{visscher201710}. 
In these biomedical applications, false negative control is pressing but fundamentally difficult. 
   
\begin{figure}[!h]
	\centering
	\includegraphics[height = 0.33\textwidth]{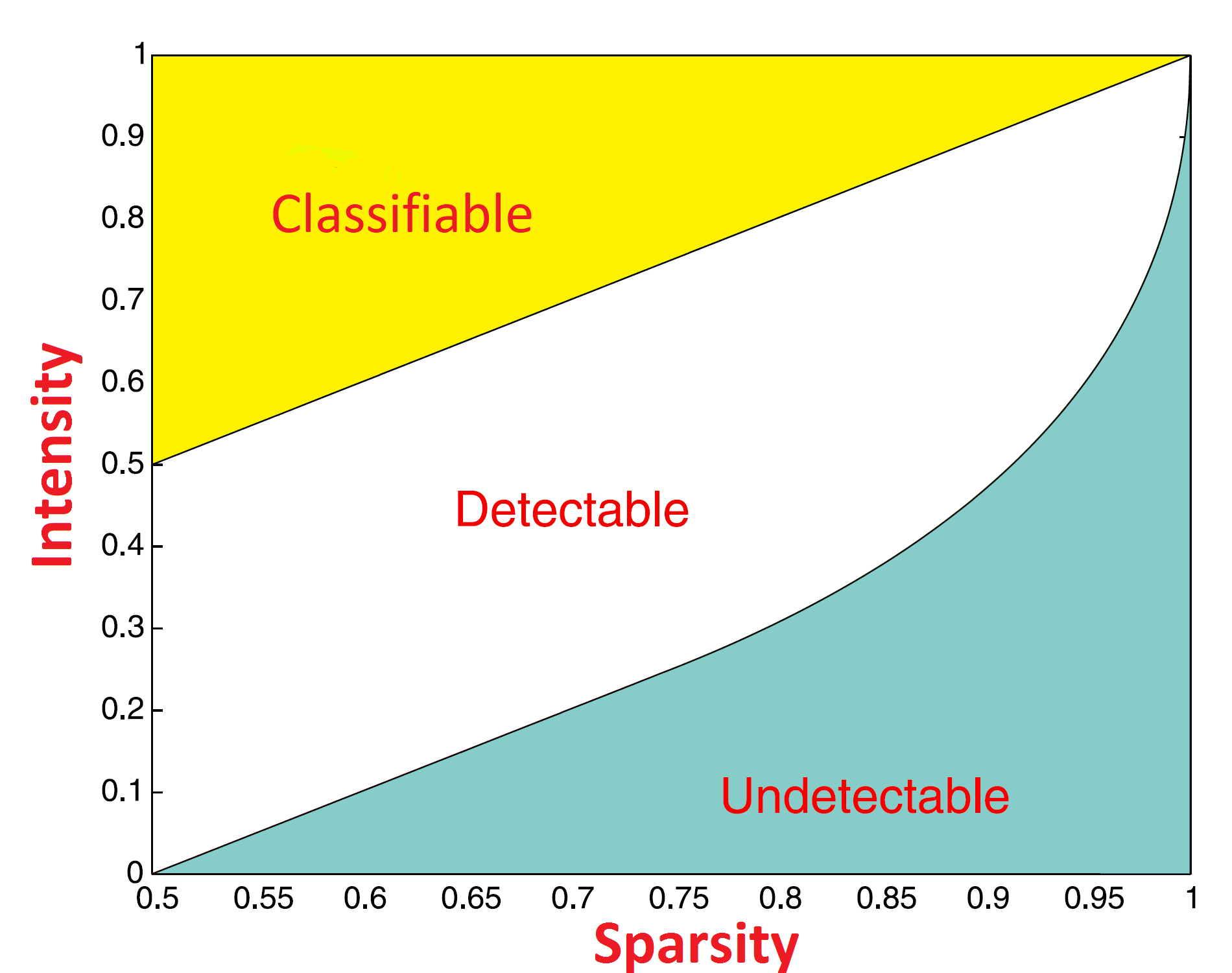}
	\caption{Phase diagram for signal detection
		and classification. Signals in the middle region are only detectable for their existence but cannot be well-separated from noise.} \label{fig:phase}
\end{figure}

In the past decade, significant contributions have been made towards two related problems in high-dimensional sparse inference. One is the detection of mixture models, which addresses the problem of ``detecting" the existence of sparse signals without specifying their exact locations (see, e.g. \cite{donoho2004higher}, \cite{arias2011global}, \cite{tony2011optimal}). The second is the separation of signals from noise variables, for which multiple testing has been used to control inflated false positive errors under high-dimensionality.  
Figure \ref{fig:phase} modified from \cite{cai2017large} illustrates the theoretical demarcation on the difficulty levels of these two problems in the setting where all variables are independent and signals are relatively sparse compared to the noise variables. Given a sparsity level, signal intensity needs to be large enough (above the undetectable region) for the signals to be detectable by a global testing procedure. For the signals to be well-separated from noise variables with negligible classification errors, signal intensity needs to be even larger (entering the classifiable region). Other versions of Figure \ref{fig:phase} can be found in \cite{donoho2004higher} and \cite{donoho2015special}. Related works analyzing phase diagrams for sparse signals under different model settings include but not limited to \cite{ji2012ups}, \cite{ji2014rate}, \cite{jin2017phase}, \cite{chen2019two}, etc. 

Figure \ref{fig:phase} also shows that signals in the middle white region can only be detected for their existence by global testing but are not separable from noise by multiple testing. Such weak signals commonly exist in large-scale data analysis and play vital roles in biomedical applications. For example, such signals can be the common genetic variants that have small effect sizes but account for a large faction of the missing heritability. In brain imaging, such weak signals often exist at the boundary of functional areas and cause difficulties for presurgical mapping and planning. 
Despite its importance, how to effectively retain detectable yet unclassifiable signals raises a bottleneck challenge in high-dimensional sparse inference. New developments are needed to bridge the methodological gap.

In this paper, we propose a new methodology framework to regulate false negative errors under measures tailored towards modern applications with high-dimensional data. More specifically, the method controls the false negative proportion (FNP $=$ number of false negatives/number of true signals)  at a user-specified level and, at the same time, regulates the amount of unnecessary false positives to achieve the FNP level.
We have found that, in real data analyses, researchers often
have different tolerance levels for false negative errors depending on their scientific inquiries. Our procedure estimates and controls FNP at a user-specified level under conditions weak enough to allow for unclassifiable signals. 
Moreover, its adaptivity to a given FNP level, such as $0.1$, allows the method to exclude a certain amount of the weakest signals to reduce possibly a large number of false positives.

The new method is developed in a realistic setting with arbitrary covariance dependence between variables.
We propose to calibrate the overall dependence strength through a parameter whose scale is compatible with the parameter space of the phase diagram presented in Figure \ref{fig:phase}. Utilizing the new calibration, we are able to asymptotically explicate the joint effect of covariance dependence, signal sparsity, and signal intensity on the proposed method. We interpret the results using a new phase diagram, which shows the power gain of the new method in retaining weak signals. 

Controlling FNP is related to some existing concepts in multiple testing literature. For example, the false non-discovery rate is defined as the expected value of the false non-discovery proportion, which represents 
the proportion of false negatives among the accepted null hypotheses for a single-step multiple testing procedure \citep{genovese2002operating, sarkar2006false}. 
In \cite{cai2016optimal}, the missed discovery rate is defined as the expected value of FNP and studied under the independence assumption. Different from the existing analyses for false non-discovery rate (FNR) and missed discovery rate (MDR),  we focus on the more challenging problem of controlling FNP, which is a random variable. Moreover, our analyses are conducted assuming arbitrary covariance dependence. As a trade-off, our theoretical results are asymptotic, not exact.  

Finite sample performance of the proposed method is studied in simulation examples where the data are generated under several commonly observed dependence structures. These dependence structures range from weak to strong in our dependence calibration system.  
The new method outperforms existing methods in adapting to a user-specified FNP control level. 
We apply the new method to analyze fMRI data from the saccade experiment in the Individual Brain Charting project \citep{pinho2018individual}. 
Its results are compared with those of other methods and with regions that are known to be associated with saccade. The new method seems to benefit from its false negative control efficiency and exhibits a nice balance in identifying functionally relevant regions and avoiding excessive noise voxels.



\section{Method and Theory} \label{sec:method}

We consider a model with continuous null distribution $F_0$. 
Suppose that the observed test statistics
\[
X_j \sim F_0 \cdot1\{j\in I_0\}+ F_j \cdot1\{j\in I_1\}, \qquad j=1,...,p,
\]
where $I_0$ is the set of indices for noise variables, $I_1$ is the set of indices for signal variables, and $F_j$ are the signal distributions.
All $I_0$, $I_1$, and $F_j$ are unknown. One can perform inverse normal transformation as $Z_j = \Phi^{-1}(F_0(X_j))$, where $\Phi^{-1}$ is the inverse of the standard normal cumulative distribution. Then, we have 
\begin{equation} \label{def:model}
	Z_j \sim \Phi \cdot1\{j\in I_0\}+ G_j \cdot1\{j\in I_1\}, \qquad j=1,...,p,
\end{equation}   
where $G_j$ are the unknown signal distributions after the transformation. For presentation simplicity,  we assume that 
$G_j(t) < \Phi(t)$ for all $t \in \mathbb{R}$. i.e., signal variables tend to show larger values than noise variables. This assumption can be easily generalized to signals with two-sided effects. 

Consider a selection rule with threshold $t$. Define the numbers of selected cases, false positives, and false negatives as
\[
\mbox{R}(t)=\sum_{j=1}^p1_{\{Z_j>t\}}, \quad \mbox{FP}(t)=\sum_{j\in I_0}1_{\{Z_j>t\}}, \quad  \mbox{FN}(t)=\sum_{j\in I_1}1_{\{Z_j\leq t\}}.
\]
The above quantities and other classification measurements including the numbers of true positives (TP) and true negatives (TN) are summarized in Table \ref{tab:confusion}.

\begin{table}[!h]
	\small
	\centering
		\begin{tabular}{l|c c|c}
			\hline
			& Not Selected & Selected & Total\\
			\hline
			Negative & TN($t$) & FP($t$) & $p-s$\\
			Positive & FN($t$) & TP($t$) & $s$\\
			\hline
			Total & $p-R(t)$ & R$(t)$ & $p$\\
			\hline
		\end{tabular} 
		\caption{Classification matrix of the selection rule with threshold $t$.} \label{tab:confusion}
\end{table}
Note that R$(t)$ and $p$ can be directly observed. FP$(t)$, FN$(t)$, TP$(t)$, TN$(t)$, and $s (= |I_1|)$ are unknown because $I_0$ and $I_1$ are unknown. 
A generalization to two-sided signal effects can be accommodated by replacing $Z_j$ with $|Z_j|$ and only allowing $t>0$.
  
Next, define FNP with respect to $t$ as 
\begin{equation} \label{def:FNP(t)}
	\mbox{FNP}(t)=\mbox{FN}(t)/s.
\end{equation}
FNP$(t)$ may be regarded as the empirical type II error that is non-decreasing with respect to $t$. 
Our new method aims to determine a selection threshold $\hat t$ such that the corresponding FNP($\hat t$) can be controlled at a desirable low level. 

Our new method is based on the approximation of FNP$(t)$ for a given $t$. Utilizing the fact that $s = \mbox{FN}(t)+ \mbox{TP}(t)$ and R$(t) = \mbox{FP}(t) + \mbox{TP}(t)$ as shown in Table \ref{tab:confusion}, we have
\begin{equation} \label{def:FNP}
\mbox{FNP}(t)= \mbox{FN}(t)/s= 1- (\mbox{R}(t)-\mbox{FP}(t))/s,
\end{equation}
where R$(t)$ is directly observable from the data. Because the noise distribution of $Z_j$ is $N(0,1)$ and there are $p-s$ noise variables, FP$(t)$ can be approximated by its mean value $E(\mbox{FP}(t)) = (p-s) \bar \Phi(t)$, where $\bar{\Phi}(t)=1- \Phi(t)$. For illustration purpose, we first assume that the true value of $s$ is known and define 
\begin{equation} \label{def:hatFNP}
\widehat{\mbox{FNP}}(t)= \max\{1-\mbox{R}(t)/s+ (p-s) \bar{\Phi}(t)/s, ~0\}.
\end{equation}
$\widehat{\mbox{FNP}}(t)$ with an estimated $s$ will be discussed in \autoref{sec:estimatedS}. 
Note that $\widehat{\mbox{FNP}}(t)$ is not the mean value of $\mbox{FNP}(t)$, and its construction does not require information of the signal distributions $G_j, j \in I_1$. If two-sided signal effects are under consideration, we can simply modify $\widehat{\mbox{FNP}}(t)$ by replacing $(p-s)$ with $2(p-s)$. 

Now given a user-specified control level, $\beta (>0)$, on FNP, we determine the selection threshold as 
\begin{equation} \label{def:hat_t}
\hat{t}(\beta)=\sup{\{t:\widehat{\mbox{FNP}}(t) < \beta\}},
\end{equation}
and select all the candidates with $Z_j > \hat t (\beta)$. If two-sided signal effects are considered, all candidates with $|Z_j| > \hat t (\beta)$ will be selected. We refer to this procedure as a dual control method. 

The proposed dual control method has several key features. 
(1) When all variables are ranked such that $Z_{(1)} \ge Z_{(2)} \ge \ldots \ge Z_{(p)}$, the dual control method selects the smallest subset from the top whose estimated FNP is less than $\beta$. Such selection process can effectively retain a high proportion of true signals while excluding unnecessary noise variables. 
(2) The proposed method can be applied as long as the marginal distribution of the noise variables is given. Its implementation does not require any information of the signal distributions nor relationships among the variables. 
(3) In real studies, researchers may have different tolerance levels for false negative errors. Adapting to a user-specified control level on FNP can substantially increase the applicability of the method. 
(4) From a practical standpoint, allowing for a certain percentage of FNP may avoid possibly a large number of false positives as a trade off. 

\subsection{Theoretical properties under the Gaussian assumption} \label{sec:FNPcontrol}

In this section, we study the asymptotic behavior of the proposed dual control method in the setting of sparse inference where signal proportion $\pi (= s/p$) converges to zero as $p \to \infty$. Specifically, we assume
\begin{equation} \label{def:proportion}
\pi = \pi_p = p^{-\gamma}, \qquad \gamma \in (0, 1).
\end{equation}
Consequently, the number of signals $s (= p^{1-\gamma})$ increases with $p$ but of a smaller order.
Such calibration on sparsity has been widely adopted in literature, see e.g. \cite{donoho2004higher}, \cite{tony2011optimal}, \cite{arias2011global}, \cite{ji2012ups}, \cite{ji2014rate}, \cite{jin2017phase}.

In this section, we explicate the combined effects of signal sparsity, signal intensity, and dependence between variables under the assumption that
\begin{equation} \label{def:multiNorm}
(Z_1, \ldots, Z_p) \sim N_p(\mu, \mathbf{\Sigma}),
\end{equation}
where $\mu$ is a $p$-dimensional vector with $\mu_j= A_j \cdot 1\{j\in I_1\}$, $A_j > 0$, and $\mathbf{\Sigma}$ is an arbitrary correlation matrix. 

Existing studies in sparse inference often assume independence between variables. For example, results in Figure \ref{fig:phase} were derived assuming normality and independence \citep{donoho2004higher, donoho2015special, cai2017large}. Here, we are interested in the performance of the new method under arbitrary covariance dependence. To this end, we calibrate the dependence effect through the following procedure: 

(1) define 
\[
\bar{\rho}=\lVert\mathbf{\Sigma}\rVert_1/p^2, \qquad \text{where~} \lVert\mathbf{\Sigma}\rVert_1 = \sum_{ij} |\sigma_{ij}|;
\]

and (2) introduce parameter $\eta$ such that
\begin{equation} \label{def:eta}
\bar \rho = \bar \rho_p = p^{-\eta} , \qquad \eta \in [0, 1].
\end{equation}
In can be seen that $\bar \rho$ summarizes the overall covariance dependence by calculating the mean absolute correlation. 
In high-dimensional data analysis with large $p$, $\bar \rho$ is often very close to zero because not every variable is correlated to all the other variables. For example, $\bar \rho$ of the $\mathbf{\Sigma}_{p \times p}$ from an autoregressive model has the order of $p^{-1}$. In order to  calibrate the dependence effect in a proper scale, the parameter $\eta$ is introduced.  Clearly, $\eta$ is in a constant scale and decreases with $\bar \rho$. $\eta=0$ corresponds to the extremely dependent case where every variable is correlated to all the other variables, and, at the opposite end, $\eta=1$ corresponds to the independent case.

With $\gamma$ and $\eta$ representing signal sparsity and overall strength of covariance dependence, respectively, we discover a lower bound condition on the signal intensity ($A_j$) for the success of the new method. The lower bound is defined as
\begin{equation} \label{def:t_min}
\mu_{min}= \min\{\mu_1, ~\mu_2\}, 
\end{equation}
where 
\[
\mu_1 = \sqrt{2\gamma \log{p}} \qquad \mbox{and} \qquad \mu_2 = \sqrt{(4\gamma-2\eta)_{+} \log{p} + 4 \log \log p}. 
\]
The lower bound $\mu_{min}$ increases with $p$ at the order of $\sqrt{\log p}$. Such order is frequently required for signal intensity level in high-dimensional sparse inference (see, e.g., \cite{ingster1994minimax}, \cite{donoho2004higher}, \cite{tony2011optimal}, \cite{arias2011global}, and \cite{cai2017large}). 
The lower bound $\mu_{min}$ that we discovered takes the value of either $\mu_1$ or $\mu_2$, depending on whichever is smaller under dependence. Specifically, $\mu_2<\mu_1$ when $\eta$ is large enough or, in other words, when covariance dependence is weak enough. As dependence gets stronger and $\eta$ gets smaller, $\mu_1< \mu_2$ and the lower bound equals to $\mu_1$ and stops to change with $\eta$. The term $\log\log p$ is a technical term for asymptotic analysis. 
We have the following theorem showing the asymptotic performance of the new method when the lower bound condition is satisfied.

\begin{theorem}\label{thm:FNPcontrol} 
	Consider model (\ref{def:multiNorm}) and a user-specified control level $\beta$ of FNP. Let $A_{min} = \min\{A_j, j\in I_1\}$ and assume $A_{min}- \mu_{min} \to \infty$ as $p\to \infty$, where $\mu_{min}$ is defined in (\ref{def:t_min}). Then the dual control method with threshold $\hat{t}(\beta)$ defined in (\ref{def:hat_t}) efficiently controls the true FNP at the level of $\beta$, i.e.,
	\begin{equation} \label{eq:FNPcontrol_1}
	P(\mbox{FNP}(\hat{t}(\beta)) \le \beta)\rightarrow 1,
	\end{equation}
	and, for any smaller set of variables associated with threshold $\tilde{t} > \hat{t}(\beta)$, 
	\begin{equation} \label{eq:FNPcontrol_2}
	P\{\mbox{FNP}(\tilde{t})>\beta -\delta) \rightarrow 1 
	\end{equation}
	for arbitrarily small constant $\delta>0$. 
\end{theorem}

\autoref{thm:FNPcontrol} shows efficiency of the new method in selecting the smallest subset of variables to achieve FNP control at the $\beta$ level. In other words, the method also controls unnecessary false positives to achieve the user-specified FNP control.  
The condition on $A_{min}$ explicates the joint effect of signal intensity, signal sparsity, and  covariance dependence on the dual control method. 

\begin{figure}[!h]
	\centering
	\includegraphics[height = 0.5\textwidth, width=0.5\textwidth]{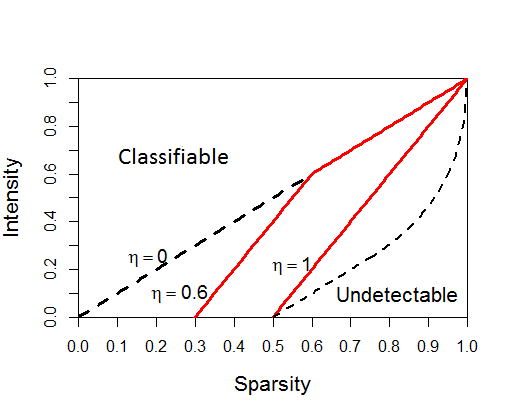}
	\caption{Dual control region under covariance dependence. Signals in the area above a solid red line can be retained by the dual control method at a pre-specified FNP level. The area increases with $\eta$ as the dependence get weaker. } \label{fig:phase_fnp}
\end{figure}

The calibration of covariance dependence through $\eta$ and the results in Theorem \ref{thm:FNPcontrol} allow us to present a new phase diagram illustrating the performance of the dual control method. More specifically, let $A_{min} = \sqrt{2 r \log p}$, $r > 0$. The condition on $A_{min}$ in \autoref{thm:FNPcontrol} can be transformed to $r > \min\{\gamma, ~2\gamma - \eta\}$ and demonstrated in a two-dimensional phase diagram with $\gamma$ as the x axis and $r$ as the y axis. The lower bound  $\min\{\gamma, ~2\gamma - \eta\}$  is illustrated in \autoref{fig:phase_fnp} by the red solid line that moves with the dependence parameter $\eta$. Signals in the area above the red solid line can be efficiently retained by the dual control method at a pre-specified level. Recall \autoref{fig:phase}, which shows that under independence, signals in the region between the two dashed lines are only detectable for their existence but not classifiable. Our result in the special case of independence ($\eta =1$) shows that many of such weak signals can be efficiently retained by the new method.




	\subsection{FNP control with estimated number of signals} \label{sec:estimatedS}
	
	Implementation of the new method requires information of the number of signals ($s$) which, in real applications, is often unknown. Existing studies for the estimation of $s$ usually assume independence among variables \citep{GW04, MR06, cai2007estimation, cai2010optimal}, and most of them are for relatively dense signals. We are interested in finding an estimator of $s$ that works in our setting with arbitrary covariance dependence. Recent study in \cite{jeng2021estimating} introduces an estimator for the signal proportion $\pi (= s/p)$ with the form $\hat \pi_{} = \max\{\hat \pi_\delta, \delta \in \Delta\}$, where $\Delta$ is a set of functions that render the most powerful consistent estimators 
	in a family of estimators under different dependence scenarios. More specifically, the family of estimators are defined as 
	\begin{equation} \label{def:MR}
		\hat{\pi}_{\delta}=\sup_{t > 0}{\frac{ R(t)/p-2\bar{\Phi}(t)-c_{p, \delta}\delta \left( t\right) }{1-2\bar{\Phi}(t)}}, 
	\end{equation}
	where $\delta(t)$ is a strictly positive function on $(0, \infty)$ and
	$c_{p, \delta}$ is the bounding sequence corresponding to $\delta(t)$. It has been discovered that  $\delta(t) = [\bar \Phi(t)]^{1/2}$ and $\delta(t) =\bar \Phi(t)$ result in powerful $\hat{\pi}_{\delta}$ under different dependence scenarios, and a new estimator is developed as 
	\begin{equation} \label{def:hat_pi}
		\hat \pi_{} = \max\{\hat \pi_{0.5}, \hat \pi_{1}\},
	\end{equation}
	where $\hat \pi_{0.5}$ denotes $\hat \pi_{\delta}$ with $\delta = [\bar \Phi(t)]^{1/2}$, and $\hat \pi_{1}$ denotes $\hat \pi_{\delta}$ with $\delta = \bar \Phi(t)$. Details about the selection of  $\delta(t)$, the construction of $c_{p, \delta}$, and the properties of  $\hat{\pi}_{\delta}$ and $\hat \pi_{}$ can be found in \cite{jeng2021estimating}. 
	
	We extend the results in \cite{jeng2021estimating} to our setting with $\gamma$ and $\eta$ representing signal sparsity and overall dependence strength, respectively, so that the conditions for the consistency of  $\hat \pi_{}$ and the conditions for the FNP control of our proposed method can be unified. Because the theoretical analysis in \cite{jeng2021estimating} is conducted on a discretized version ($\hat \pi_{\delta}^*$) of $\hat \pi_{\delta}$, which is defined by replacing $\sup_{t>0}$ in (\ref{def:MR}) with $\max_{t \in \mathbb{T}}$, where $\mathbb{T} = [1, \sqrt{5 \log p}] \cap \mathbb{N}$, we present the unified conditions for the discretized version of  $\hat \pi_{}$ as well and define it as  $\hat \pi_{}^* = \max\{\hat \pi^*_{0.5}, \hat \pi^*_{1}\}$.

	\begin{proposition} \label{prop:estimated_s}
		Assume the same conditions as in Theorem \ref{thm:FNPcontrol}.
		Then, for any constant $\epsilon>0$,  
		\begin{equation} \label{eq:hat_pi}
			P((1-\epsilon) < \hat \pi_{}^*/\pi  < 1)\rightarrow 1.
		\end{equation} 
		Replace $s$ in (\ref{def:hat_t}) with the estimator $\hat s = p \cdot \hat \pi_{}^*$ and denote the selection threshold as $\hat t_{\hat s}(\beta)$. We have
		\begin{equation} \label{eq:FNPcontrol_1s}
			P(\mbox{FNP}(\hat{t}_{\hat s}(\beta)) \le \beta)\rightarrow 1
		\end{equation}
		and, for any threshold $\tilde{t} > \hat{t}_{\hat s}(\beta)$, 
		\begin{equation} \label{eq:FNPcontrol_2s}
			P\{\mbox{FNP}(\Tilde{t})>\beta -\delta) \rightarrow 1 
		\end{equation}
		for arbitrarily small constant $\delta>0$. 	
	\end{proposition}
	
	Proposition \ref{prop:estimated_s} shows that the same set of conditions are sufficient for both the consistency of $\hat \pi_{}^*$ and the efficient FNP control of the dual control method implemented with $\hat \pi_{}^*$. This result can also be interpreted by the phase diagram in \autoref{fig:phase_fnp}, where the solid red line that moves with the dependence level $\eta$ now indicates the required signal intensity for both $s$ estimation and FNP control. In other words, for signals above the solid red line, although they may not be individually identifiable, 
	we can still perform meaningful inference by not only estimating their total numbers but also effectively retaining them at a target FNP level.   
	
	Generally speaking, if we want to study the power of some method, a condition on the signal intensity is unavoidable. When signal intensity is too low, no methods can even detect their existence as illustrated in Figure \ref{fig:phase}, let along the more challenging tasks of FNP control. 
	When the condition on signal intensity does not hold, the estimator $\hat \pi_{}$ tends to underestimate  because of its lower bound property \citep{jeng2021estimating}. Consequently, the dual control method implemented with $\hat s = p \cdot \hat{\pi}_{}$ tends to have a threshold higher than the ideal threshold that achieves the target level of $\beta$, which results in conservative variable selection with less false positives but inflated FNP. This tendency is observed in simulation examples with low signal-to-noise ratio in \autoref{sec:simulation_estS}.

\subsection{Computational algorithm}
	
We provide algorithms to calculate the bounding sequence $c_{p, \delta}$, to derive the proportion estimator $\hat \pi$ in (\ref{def:MR}), and to select variables by the dual control method.

	\begin{algorithm}[]
		\caption{Bounding Sequence $c_{p, \delta}$}\label{alg_1}	
		\begin{algorithmic}[1]
			\Statex {\bf Input:} N sets of $\{w_1, \ldots, w_p\}$ generated by the joint null distribution of $\{z_1, \ldots z_p\}$ 
			\Statex {\bf Output:} bounding sequences $c_{p, 0.5}$ and $c_{p, 1}$
			\State {\bf for} $a=1, 2\ldots, N$ {\bf do} 
			\Statex Rank the $a$-th set of $\{w_1, \ldots, w_p\}$ such that $w_{(1)} > w_{(2)} > \ldots > w_{(p)}$
			\Statex Compute 
			\[
			V_{0.5, a} = \max_{1 \le j \le p} {|j / p - \bar \Phi(w_{(j)})| \over \sqrt{\bar \Phi(w_{(j)})}} \qquad \mbox{and} \qquad V_{1, a} = \max_{1 \le j \le p} {|j / p - \bar \Phi(w_{(j)})| \over \bar \Phi(w_{(j)})} 
			\]
			\State {\bf end for}
			\State Compute $c_{p, 0.5}$ and $c_{p, 1}$ as the $(1-1/\sqrt{\log p})$-th quantiles of the empirical distributions of $V_{0.5, a}$ and $V_{1, a}$, respectively
		\end{algorithmic}
	\end{algorithm}
	
	Note that the computation of  $c_{p, \delta}$ in Algorithm \ref{alg_1} requires input of statistics generated by the joint null distribution. When the joint null distribution is unknown in  real applications, we often can simulate such statistics non-parametrically. 
	For example, when$\{z_1, \ldots z_p\}$ are a set of test statistics for associations between a set of explanatory variables and a response variable, we may randomly shuffling only the sample of the response variable to remove the potential associations and then calculate the test statistics. More details for such permutation approaches can be found in \cite{westfall1993resampling}. By this way, we can generate $N$ sets of test statistics, where $N$ is a predetermined large number, such as 1000.

	\begin{algorithm}[H]
		\caption{Signal Proportion Estimator}\label{alg}	
		\begin{algorithmic}[1]
			\Statex {\bf Input:} test statistics $\{z_1, \ldots z_p\}$ and bounding sequences $c_{p, 0.5}$ and $c_{p, 1}$
			\Statex {\bf Output:} a proportion estimate $\hat \pi$
			\State Rank the variables by their test statistics so that $z_{(1)} > z_{(2)} > \ldots > z_{(p)}$	
			\State Compute 
			\[
			\hat{\pi}_{0.5}= \max_{1 \le j \le p}{\frac{ j/p-\bar{\Phi}(z_{(j)})-c_{p,0.5} \sqrt{\bar \Phi(z_{(j)})}} {1-\bar{\Phi}(z_{(j)})}} \qquad \mbox{and} \qquad \hat{\pi}_{1}= \max_{1 \le j \le p}{\frac{ j/p-\bar{\Phi}(z_{(j)})-c_{p,1} \bar \Phi(z_{(j)})} {1-\bar{\Phi}(z_{(j)})}}
			\]
			\State Obtain $\hat \pi = \max\{\hat{\pi}_{0.5}, \hat{\pi}_{1}\}$
		\end{algorithmic}
	\end{algorithm}

	\begin{algorithm}[H]
		\caption{Dual Control Procedure }\label{alg}	
		\begin{algorithmic}[1]	
			\Statex {\bf Input:} test statistics $z_1, \ldots z_p$, a user-specified $\beta$, and a proportion estimate $\hat \pi$
			\Statex {\bf Output:} a set of selected variables		
			\State Rank the variables by their test statistics so that $z_{(1)} > z_{(2)} > \ldots > z_{(p)}$		
			\State Compute $\hat s = \hat \pi p$
			\State For $j=1, 2, \ldots, p$, compute
			\[
			\widehat{FNP}(z_{(j)})= \max\{1 - j/\hat s + (p-\hat s) \bar \Phi(z_{(j)})/\hat s, ~0\}
			\]		
			\State Let $k=1$
			\Statex {\bf while}  $ \widehat{FNP}(z_{(k)}) 
			\ge \beta$ {\bf do}  $k = k+1$
			\Statex {\bf end while} 	
			\State Obtain variables ranked at $1, \ldots, k-1$		
		\end{algorithmic}
	\end{algorithm}
	

\section{Simulation}  \label{sec:simulation}
	
We provide simulation examples to demonstrate the finite-sample performance of the dual control method. In these examples,  a series of test statistics are generated as
	\begin{equation} \label{eq:Z}
	(Z_1, \ldots, Z_p)^T \sim N((\mu_1, \ldots, \mu_p)^T, \mathbf{\Sigma}),
	\end{equation}
	where $p=2000$, $\mu_j = A \cdot 1\{j \in I_1\}$, and $I_1$ is a set of indices randomly sampled from $\{1,..., p\}$ with cardinality $s = |I_1| = p^{1-\gamma}$.  We consider three different dependence structures:
	\begin{itemize}
		\item Model 1 [Autoregressive]: $\mathbf{\Sigma}=(\sigma^{(1)}_{ij})$, where $\sigma^{(1)}_{ij}=\lambda^{|i-j|}$ for $1\leq i,j\leq p$.
		\item Model 2 [Block dependence]: $\mathbf{\Sigma}=\mathbf{I}_{p/k}\otimes\mathbf{D}$, where $\mathbf{D}$ is a $k\times k$ matrix with diagonal entries 1 and off-diagonal entries $r$.
		\item Model 3 [Factor model]: $\mathbf{\Sigma}=(\sigma^{(3)}_{ij})$, where $\sigma^{(3)}_{ij}=V_{ij}/\sqrt{V_{ii}V_{jj}}$ for $1\leq i,j\leq p$, $\mathbf{V}=\tau\mathbf{h}\mathbf{h}^T+\mathbf{I}_p$ with $\tau \in (0, 1)$ and $\mathbf{h}\sim N(\mathbf{0}, \mathbf{I}_p)$.
	\end{itemize}

For dependence parameters, we have $\lambda=0.2, k=40, r=0.5$, and $\tau=0.5$, so that the dependence parameter $\eta$ decreases from 0.95 in Model 1, to 0.57 in Model 2, to 0.23 in Model 3. In our calibration system, dependence increases from very weak in Model 1, to moderately strong in Model 2, to very strong in Model 3. 

\subsection{Comparison with multiple testing methods}	\label{sec:sim_multiple}

We compare the proposed method with the classical BH-FDR procedure \citep{benjamini1995}. Specifically, we show that the FNP control property of our method cannot be achieved by manipulating the nominal level of BH-FDR. 

In this section, we set $\gamma = 0.3$, which corresponds to $s=205$, i.e. there are 205 signal variables with elevated mean values randomly located among 1795 noise variables. The signal intensity $A=3$ and $2$. 

The proposed method is denoted as DCOE (acronym for Dual Control Of Errors). In \autoref{tab:compare}, we summarize the realized FNP of DCOE with its selection threshold constructed as in (\ref{def:hat_t}) from 100 replications. The results are compared with those of BH-FDR. 
We also compare the realized false discovery proportion (FDP) of the two methods. Both methods are applied with varying nominal levels. 

	\begin{table}[!h]
		\caption{Mean values and standard deviations (in brackets) of the realized FNP and FDP of the proposed DCOE method and the existing BH-FDR procedure. 
		} \label{tab:compare}
		\centering
		\setlength{\tabcolsep}{1.5mm}{
			\begin{tabular}{ll l| c c }
				& & & FNP & FDP   \\
				\hline
				A=3 & Autoregressive & DCOE(${\bf \beta= 0.2}$) & {\bf 0.198} (0.023) &  0.149 (0.035) \\
				& & DCOE($\beta={\bf 0.1}$) & {\bf 0.101} (0.037)  & 0.307 (0.084) \\
				& & BH-FDR($\alpha={\bf 0.05}$) & 0.378 (0.040)  & {\bf 0.044} (0.018) \\
				& & BH-FDR($\alpha={\bf 0.2}$) & 0.166 (0.028) & {\bf 0.183} (0.028) \\
				\cline{2-5} 		
				& Block dependence & DCOE($\beta={\bf 0.2}$) & {\bf 0.170} (0.086) & 0.259 (0.247) \\
				& & DCOE($\beta={\bf 0.1}$) & {\bf 0.089} (0.074) & 0.444 (0.268) \\
				& & BH-FDR($\alpha={\bf 0.05}$) & 0.387 (0.071) & {\bf 0.047} (0.036) \\
				& & BH-FDR($\alpha={\bf 0.2}$) & 0.169 (0.049)  & {\bf 0.180} (0.070) \\
				\cline{2-5}
				& Factor model & DCOE($\beta={\bf 0.2}$) & {\bf 0.160} (0.099) &  0.262 (0.215) \\
				& & DCOE($\beta={\bf 0.1}$) & {\bf 0.084} (0.104) &  0.533 (0.227) \\
				& & BH-FDR($\alpha={\bf 0.05}$) & 0.400 (0.043) & {\bf 0.041} (0.049) \\
				& & BH-FDR($\alpha={\bf 0.2}$) & 0.176 (0.049)  & {\bf 0.170} (0.113) \\
				\hline
				\hline
				A=2 & Autoregressive & DCOE(${\bf \beta= 0.2}$) & {\bf 0.201} (0.069) & 0.576  (0.073) \\
				& & DCOE($\beta={\bf 0.1}$) & {\bf 0.114} (0.069)  & 0.688 (0.085) \\
				& & BH-FDR($\alpha={\bf 0.05}$) & 0.889 (0.035)  & {\bf 0.040} (0.043) \\
				& & BH-FDR($\alpha={\bf 0.2}$) & 0.630 (0.050)  & {\bf 0.181} (0.044) \\
				\cline{2-5}
				& Block dependence & DCOE($\beta={\bf 0.2}$) & {\bf 0.193} (0.132)  & 0.607 (0.161) \\
				& & DCOE($\beta={\bf 0.1}$) & {\bf 0.132} (0.116) & 0.685 (0.152) \\
				& & BH-FDR($\alpha={\bf 0.05}$) & 0.894 (0.058)  & {\bf 0.055} (0.070) \\
				& & BH-FDR($\alpha={\bf 0.2}$) & 0.647 (0.097)  & {\bf 0.182} (0.093) \\
				\cline{2-5}
				& Factor model & DCOE($\beta={\bf 0.2}$) & {\bf 0.188} (0.169) &  0.642 (0.107) \\
				& & DCOE($\beta={\bf 0.1}$) & {\bf 0.156} (0.159) & 0.693 (0.098) \\
				& & BH-FDR($\alpha={\bf 0.05}$) & 0.886 (0.088) & {\bf 0.032} (0.061) \\
				& & BH-FDR($\alpha={\bf 0.2}$) & 0.659 (0.088)  & {\bf 0.160} (0.123) \\
				\hline
			\end{tabular}}
		\end{table}
		
		It can be seen that for Model 1 [Autoregressive], the mean values of FNP for DCOE are  fairly close to their corresponding nominal levels ($\beta=0.2$ and $0.1$ ). 
		In the more challenging cases generated by Model 2 [Block dependence] and 3 [Factor model], 
		the differences between realized FNPs of DCOE and the nominal levels increase slightly. These observations agree with the theoretical results in Section \ref{sec:FNPcontrol}. 
		
		On the other hand, \autoref{tab:compare} shows that the mean values of the realized FDP of BH-FDR are fairly close to their corresponding nominal levels of $\alpha$. 
		However, the mean values of FNP for BH-FDR are generally larger than those of DCOE. It is true that one can increase the nominal level of BH-FDR to reduce FNP, but it is unclear by how much one should increase $\alpha$ to achieve a target level of FNP. For example, suppose we have a target $\beta = 0.2$. It seems that when $A=3$, BH-FDR with $\alpha=0.2$ has the mean values of FNP in the range of $0.166-0.176$, which are close to the target $\beta$. 
		However, when $A=2$, BH-FDR with $\alpha=0.2$ has mean values of FNP in the range of $0.63 - 0.66$, which are much bigger than $\beta=0.2$.
		
		These results demonstrate the fundamental differences between the existing multiple testing procedures and the proposed dual control method as they serve for very different purposes. The proposed method aims to efficiently retain signals at a target FNP level, for which the multiple testing methods cannot achieve. On the other hand, the new method pays the price of having a higher FDP when signals are relatively weak.


		


	\subsection{Comparison with other false negative control methods} \label{sec:simulation_estS}
	In this section,  we compare the empirical performances of DCOE with existing methods that are designed to regulate false negatives. Such methods are relatively rare and have only appeared recently. 
	For example, the AFNC method in \cite{jeng2016rare} was proposed to control the so-called signal missing rate to detect rare variants in genome-wide association study \citep{jeng2016rare}; 
	the MDR method in \cite{cai2016optimal} was proposed to control the mean value of FNP using an empirical Bayesian approach \citep{cai2016optimal}; 
	the AdSMR method in \cite{jeng2019efficient} focuses on controlling a FNP-based exceedance probability; and the FNC-Reg approach in \cite{JengChen2019} considers controlling FNP in linear regression. Among these methods, AFCN, MDR, and FNC-Reg are more comparable to DCOE because they all require the input of a user-specified control level. However, AFNC and MDR were developed assuming independence between variables, and FNC-Reg considered specific dependence and signal sparsity conditions to facilitate accurate precision matrix estimation and bias mitigation in linear regression. All these methods require estimates of the number of signals. For a fair comparison, we implement the same estimator $\hat s = p \cdot \hat \pi$, where $\hat \pi$ is defined in (\ref{def:hat_pi}), to these methods. 
	
	
	
	The performances of the methods are evaluated by three measures. The first two measures, FNP and FDP are the same as in \autoref{tab:compare}.  The last measure is the Fowlkes-Mallows index \citep{fowlkes1983method, halkidi2001clustering}, which summarizes the measures of FNP and FDP by calculating the geometric mean of ($1-$FNP) and ($1-$FDP), i.e., 
	\[
	\mbox{FM-index} = \sqrt{(1-\mbox{FNP}) \times (1-\mbox{FDP})}.
	\]   
	Higher values of the FM-index indicate better classification results. The FM-index is a sensible summary measure in high-dimensional settings with sparse alternative cases because the scale of FDP is more comparable to that of FNP than the classical false positive proportion (FPP = number of false positives/number of null cases). Because the methods presented in this section all focus on false negative control, it is appropriate to use FM-index to compare their efficiency.  
	
	We generate the test statistics by (\ref{eq:Z}) with a covariance matrix that has 20 diagonal blocks with block sizes randomly generated from 10 to 100. The non-zero off-diagonal correlations are set at 0.5. The dependence parameter $\eta$ varies from sample to sample due to the random block size.

	In the first set of examples, signal sparsity is fixed with $\gamma = 0.3$, and signal intensity (A) increases from $3$ to $5$.  First, 
	because the estimator $\hat s$ underestimates the true $s$ when signal intensity is not strong enough, DCOE implemented with $\hat s$ selects less variables than actually needed to reach the nominal level of $\beta$. These result in inflated realized FNP as seen in \autoref{tab:compare_mu}. 
	Further, as $A$ increases, the mean value of FNP of DCOE gets closer to the nominal level of $\beta$, which agrees with the claims in (\ref{eq:FNPcontrol_1s}) and (\ref{eq:FNPcontrol_2s}) in \autoref{prop:estimated_s}.  
	Among the three methods presented in \autoref{tab:compare_mu}, DCOE shows a clear tendency to adapt to the nominal level of $\beta$ as $A$ increases, which is not observed for the other two methods. In terms of the FM-index, DCOE generally outperforms the other two methods especially when signal intensity gets stronger.  
	
	\begin{table}[!h]
		\caption{Effect of signal intensity on different FN control methods with estimated s. Mean values and standard deviations (in brackets) of the realized FNP, FDP, and the FM-index are presented for the proposed DCOE method and two existing methods, AFNC and MDR from 100 replications. }\label{tab:compare_mu}
		\centering
		\setlength{\tabcolsep}{1.5mm}{
			\begin{tabular}{l c| c c c}
				& &  FNP & FDP & FM-index \\
				\hline 
				$A=3$  & DCOE($\beta={\bf 0.1}$) & {\bf 0.28} (0.12) &  0.13 (0.14) & {\bf 0.78} (0.04) \\
				& AFNC($\beta={\bf 0.1}$) &  0.19 (0.11)  &  0.22 (0.18) & 0.78 (0.07) \\
				& MDR($\beta={\bf 0.1}$) &  0.09 (0.08) & 0.49 (0.26) &  0.65 (0.15)  \\
				\hline
				$A=4$ & DCOE($\beta={\bf 0.1}$)  & {\bf 0.16} (0.07) & 0.06 (0.12) &  {\bf 0.88} (0.05) \\
				& AFNC($\beta={\bf 0.1}$) & 0.07 (0.04) &  0.14 (0.18) & 0.89 (0.10) \\
				& MDR($\beta={\bf 0.1}$) &  0.06 (0.05) & 0.37 (0.32) & 0.74 (0.18) \\
				\hline
				$A=5$ & DCOE($\beta={\bf 0.1}$) &  {\bf 0.10} (0.04)  & 0.04 (0.12)  & {\bf 0.93} (0.06) \\
				& AFNC($\beta={\bf 0.1}$) & 0.01 (0.01)  & 0.36 (0.35) & 0.75 (0.26) \\
				& MDR($\beta={\bf 0.1}$) & 0.04 (0.04) & 0.36 (0.33)  & 0.75 (0.19) \\
				\hline
			\end{tabular}}
		\end{table}

		In the second set of examples, signal intensity is fixed at $5$ and signal sparsity $\gamma$ increases from 0.3 to 0.5, so that the number of signals deceases from 205 to 45. The nominal level of $\beta$ also varies. Results summarized in \autoref{tab:compare_beta} show that when signals get sparser with larger $\gamma$, the performances of all three methods deteriorate by including more noise variables. However, DCOE continues to outperform the other two methods in adapting to different nominal levels and incurring less false positives. Its advantage seems to be more prominent when signals get more sparse.  
		
		

		\begin{table}[!h]
			\caption{Effect of nominal level and signal sparsity on different FN control methods.}\label{tab:compare_beta}
			\centering
			\setlength{\tabcolsep}{1.5mm}{
				\begin{tabular}{l c|ccc}
					&  & FNP & FDP & FM-index \\
					\hline 
					$\gamma=0.3$   & DCOE($\beta={\bf 0.1}$) &  {\bf 0.10} (0.04)  & 0.04 (0.12)  & {\bf 0.93} (0.06) \\
					& AFNC($\beta={\bf 0.1}$)  & 0.01 (0.01)  & 0.36 (0.35) & 0.75 (0.26) \\
					& MDR($\beta={\bf 0.1}$) & 0.04 (0.04) & 0.36 (0.33)  & 0.75 (0.19) \\
					\cline{2-5}
					& DCOE($\beta={\bf 0.2}$)  & {\bf 0.19} (0.06) & 0.02 (0.08) & {\bf 0.89} (0.03) \\
					& AFNC($\beta={\bf 0.2}$)  & 0.01 (0.01) & 0.21 (0.29) & 0.86 (0.20) \\
					& MDR($\beta={\bf 0.2}$)  & 0.08 (0.09) & 0.26 (0.28) & 0.80 (0.14) \\
					\hline
					$\gamma=0.5$ & DCOE($\beta={\bf 0.1}$) & {\bf 0.09} (0.05) & 0.15 (0.29) & {\bf 0.85} (0.18) \\
					& AFNC($\beta={\bf 0.1}$) & 0.00 (0.01) & 0.73 (0.28) & 0.44 (0.27) \\
					& MDR($\beta={\bf 0.1}$)  & 0.03 (0.05) & 0.56 (0.44) & 0.54 (0.34) \\
					\cline{2-5}
					& DCOE($\beta={\bf 0.2}$)  & {\bf 0.16} (0.09) & 0.12 (0.27) & {\bf 0.83} (0.15) \\
					& AFNC($\beta={\bf 0.2}$)  & 0.01 (0.02) & 0.60 (0.36) & 0.55 (0.31) \\
					& MDR($\beta={\bf 0.2}$)  & 0.07 (0.10) & 0.54 (0.45) & 0.54 (0.32) 	\\
					\hline
				\end{tabular}}
			\end{table}

\subsection{Additional insight on $\mu_{min}$} \label{sec:simulation_FNP}

We provide additional insight on the lower bound $\mu_{min}$ for FNP control with finite sample. Recall that $\mu_{min} = \min\{\mu_1, \mu_2\}$, where $\mu_1$ depends on signal sparsity through $\gamma$,  $\mu_2$ depends on both signal sparsity and dependence through $\gamma$ and $\eta$. We set signal sparsity parameter $\gamma = 0.3$ and consider the three dependence models as in Section \ref{sec:sim_multiple}. 

\begin{figure}[!h]
	\centering
	\includegraphics[height = 5cm, width=4.8cm]{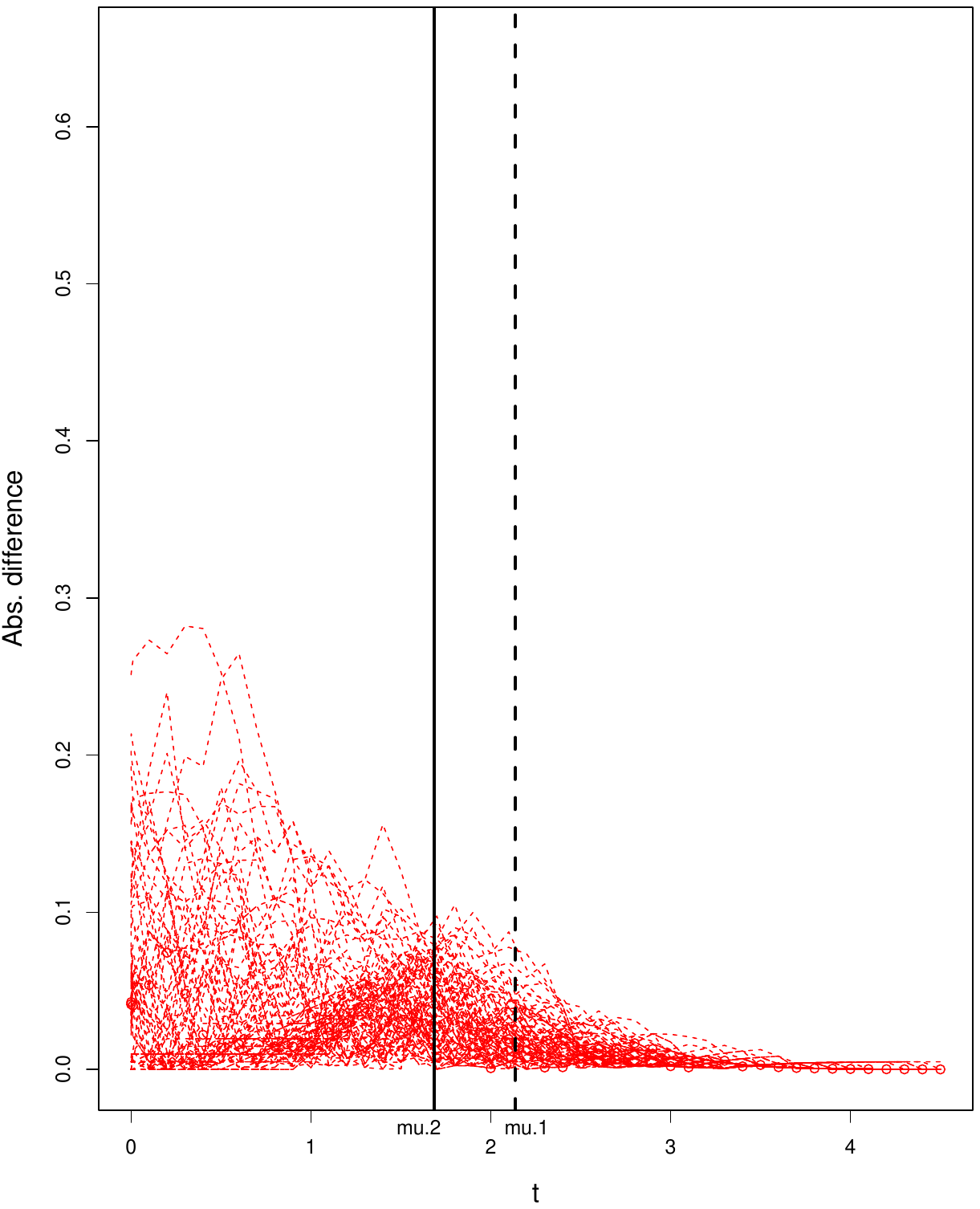}
	\includegraphics[height = 5cm, width=4.8cm]{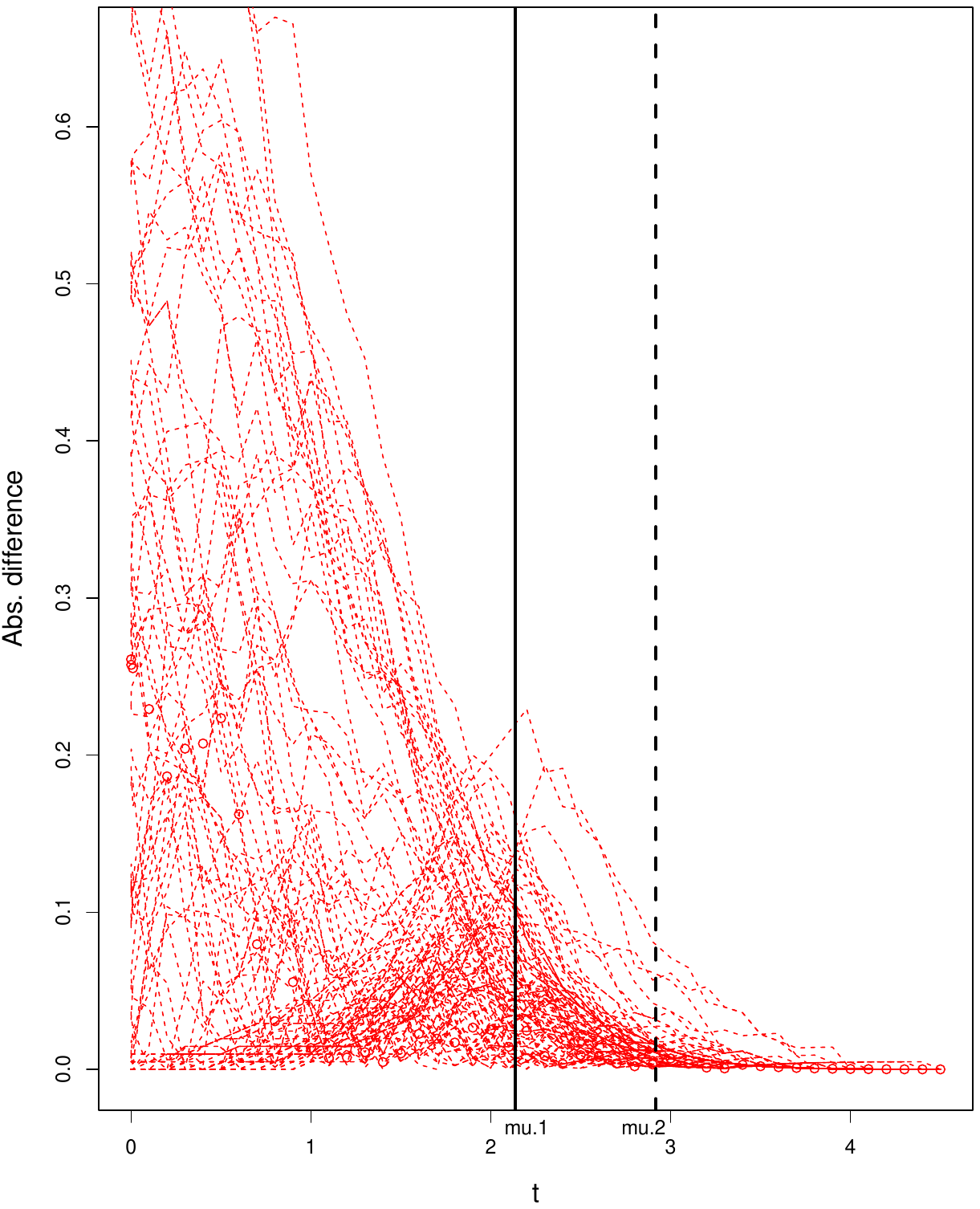}
	\includegraphics[height = 5cm, width=4.8cm]{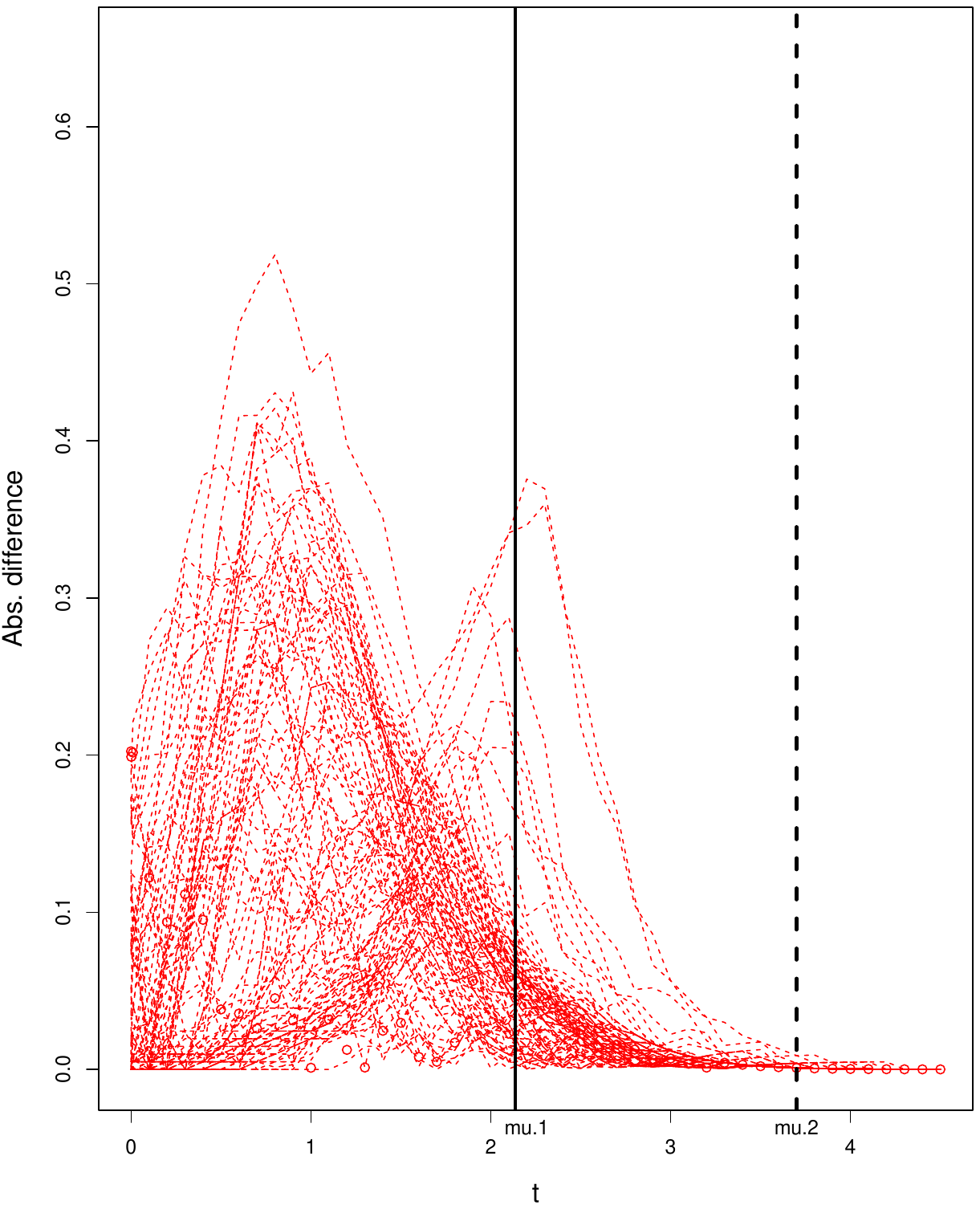}
	\caption{Estimation accuracy of $\widehat{\mbox{FNP}}(t)$ based on 100 replications. Plots from left to right are generated under Model 1 - 3. The solid vertical lines represent the boundary value $\mu_{min} = \min\{\mu_1, \mu_2\}$. Model 1 has very weak dependence with $\eta = 0.95$ and $\mu_{min}= \mu_2 = 1.69$. Model 2 has moderately strong dependence with $\eta = 0.57$ and $\mu_{min}= \mu_1 = 2.14$. Model 3 has very strong dependence with $\eta = 0.23$ and $\mu_{min}= \mu_1 = 2.14$.} \label{fig:approx}
\end{figure}

Specifically, we have $(\mu_1=2.14, \mu_2=1.69)$ for Model 1,  $(\mu_1=2.14, \mu_2=2.92)$ for Model 2, and $(\mu_1=2.14, \mu_2=3.69)$ for Model 3. Consequently, $\mu_{min} = 1.69$ for Model 1 but remains the same at $2.14$ for Model 2 and 3. These values of $\mu_{min}$ are illustrated as the solid vertical lines in \autoref{fig:approx}. The dotted vertical lines represent $\mu_1$ or $\mu_2$ whichever is larger. 
The dotted curves represent the absolute difference between the estimator $\widehat{\mbox{FNP}}(t)$ as defined in (\ref{def:hatFNP}) and the true FNP$(t)$ from 100 replications.
It can be seen  that the approximation accuracy of $\widehat{\mbox{FNP}}(t)$ increases with $t$, and the majority of the replicated differences are close to $0$ after passing $\mu_{min}$. This result hold in all three examples with very different dependence structures. 
Since the proposed method replies on the approximation accuracy of $\widehat{\mbox{FNP}}(t)$, results in Figure \ref{fig:approx} support the role of $\mu_{min}$ in the lower bound condition in Theorem \ref{thm:FNPcontrol}.

			\section{Application} \label{sec:application}
			
			We obtained the fMRI data from the Individual Brain Charting (IBC) Project, which is a publicly available  high-resolution fMRI dataset for cognitive mapping \citep{pinho2018individual}. The dataset refers to a cohort of 12 participants performing different tasks, addressing both low- and high- level cognitive functions. We focus on the data from the saccade experiment for spatial cognition, in which ocular movements were performed according to the displacement of a fixation cross from the center toward peripheral locations in the image displayed. 
			
			The data were collected using a Gradient-Echo (GE) pulse, whole-brain Multi-Band (MB) accelerated Echo-Planar Imaging (EPI) T2$^\star$-weighted sequence with Blood-Oxygenation-Level-Dependent (BOLD) contrasts, and preprocessed using {\it PyPreprocess}, a collection of python tools for preprocessing fMRI data.  In order to assess the statistical significance of the differences among evoked BOLD responses, test statistics are computed at every voxel for each contrast using General Linear Model (GLM). All images are confined to an average mask of the gray matter across subjects, which yields 371,817 voxels at the chosen resolution. More details about the dataset can be found in \cite{pinho2018individual}.   
			
			We apply the proposed method (DCOE), the popular BH-FDR procedure, and the existing AFNC and MDR methods to the statistical maps of 12 participants. The numbers of selected voxels for each participant are reported in \autoref{tab:fmri}. 
			Among all the methods, BH-FDR selects the least voxels and MDR selects the most voxels. DCOE selects more than BH-FDR and less than AFNC and MDR. 
			
			\begin{table}[h]
				\caption{Numbers of selected voxels by different methods.} \label{tab:fmri}
				\centering
				\setlength{\tabcolsep}{3mm}{
					\begin{tabular}{c|c|c|c|c}
						\hline
						Subject 
						& BH-FDR($\alpha=0.05$) & DCOE($\beta=0.1$) &
						AFNC($\beta=0.1$) & MDR($\beta=0.1$)  \\
						\hline
						\hline
						sub-1 & 18500  & 19379  & 24837  & 27793 \\
						sub-2 & 44902  & 52614  & 70981  & 78288 \\
						sub-4 & 15291  & 31852  & 44893  & 60356 \\
						sub-5 & 8623   & 23236  & 37333  & 62976 \\
						sub-6 & 17778  & 29298  & 40537  & 47812 \\
						sub-7 & 25011  & 51405  & 76931  & 88407 \\
						sub-8 & 35463  & 47467  & 65238  & 74915 \\
						sub-9 & 20846  & 35662  & 50516  & 63989 \\
						sub-11 & 28586  & 32989  & 43147 & 48892 \\
						sub-12 & 20469  & 27388  & 36256  & 38353 \\
						sub-13 & 21365  & 44094  & 63192  & 74508 \\
						sub-14 & 21067  & 58494  & 85889  & 96651 \\
						\hline
					\end{tabular}}
				\end{table}

\begin{figure}
					\begin{subfigure}{.32\textwidth}
						\centering
						\includegraphics[width=.8\linewidth]{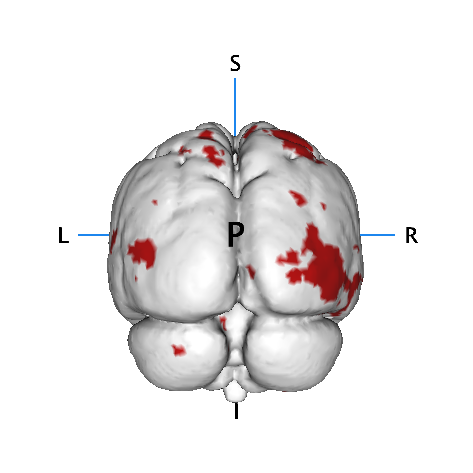}
						\caption{BH-FDR (posterior)}
						\label{fig:sfig1}
					\end{subfigure}%
					\begin{subfigure}{.32\textwidth}
						\centering
						\includegraphics[width=.8\linewidth]{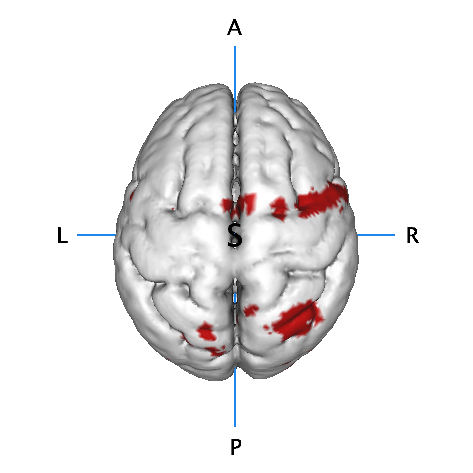}
						\caption{BH-FDR (superior)}
						\label{fig:sfig2}
					\end{subfigure}
					\begin{subfigure}{.32\textwidth}
						\centering
						\includegraphics[width=.8\linewidth]{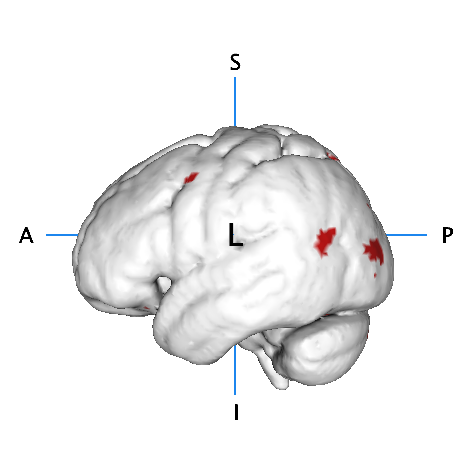}
						\caption{BH-FDR (left)}
						\label{fig:sfig3}
					\end{subfigure}
					
					\begin{subfigure}{.32\textwidth}
						\centering
						\includegraphics[width=.8\linewidth]{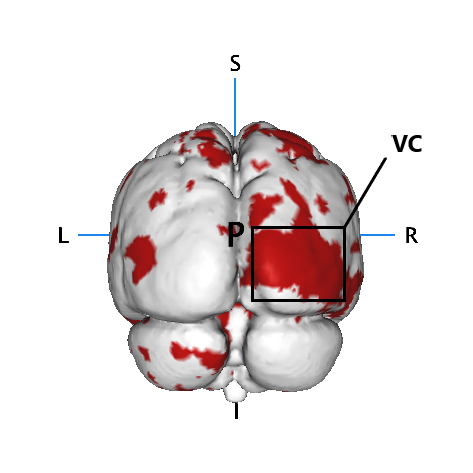}
						\caption{DCOE (posterior)}
						\label{fig:fnpc_p}
					\end{subfigure}
					\begin{subfigure}{.32\textwidth}
						\centering
						\includegraphics[width=.8\linewidth]{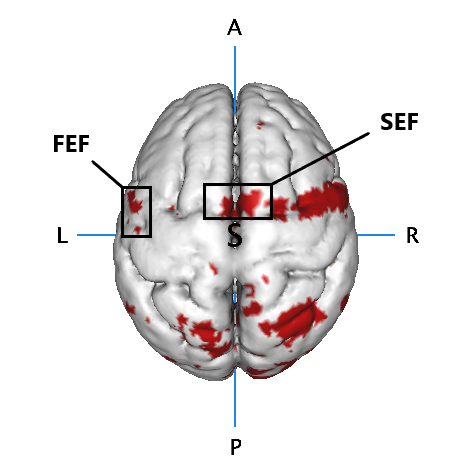}
						\caption{DCOE (superior)}
						\label{fig:fnp_s}
					\end{subfigure}
					\begin{subfigure}{.32\textwidth}
						\centering
						\includegraphics[width=.8\linewidth]{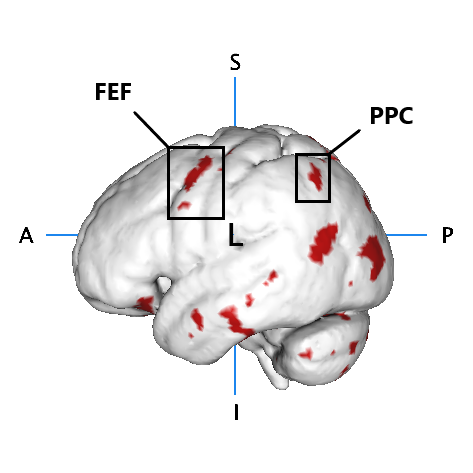}
						\caption{DCOE (left)}
						\label{fig:fnp_l}
					\end{subfigure}
					
					\begin{subfigure}{.32\textwidth}
						\centering
						\includegraphics[width=.8\linewidth]{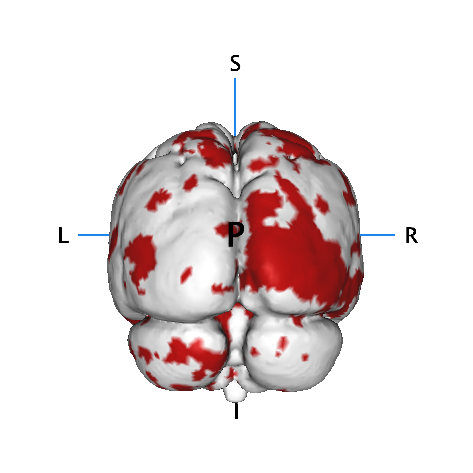}
						\caption{AFNC (posterior)}
						\label{fig:sfig7}
					\end{subfigure}
					\begin{subfigure}{.32\textwidth}
						\centering
						\includegraphics[width=.8\linewidth]{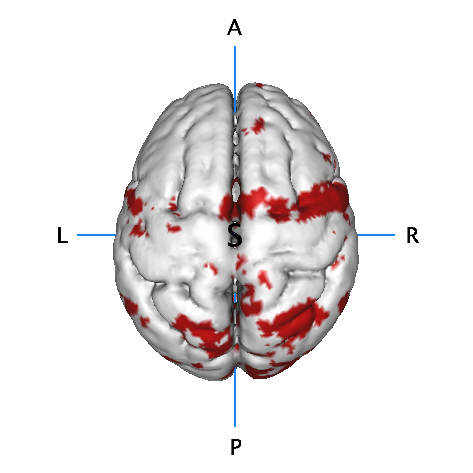}
						\caption{AFNC (superior)}
						\label{fig:sfig8}
					\end{subfigure}
					\begin{subfigure}{.32\textwidth}
						\centering
						\includegraphics[width=.8\linewidth]{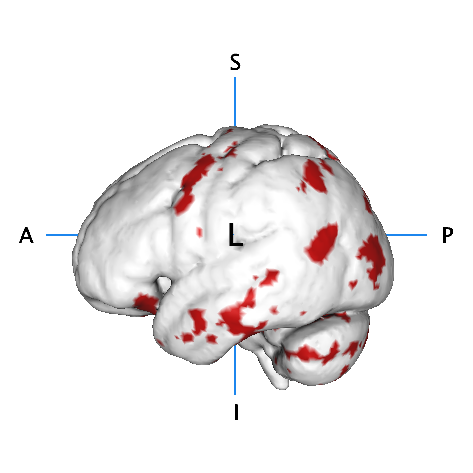}
						\caption{AFNC (left)}
						\label{fig:sfig9}
					\end{subfigure}
					
					\begin{subfigure}{.32\textwidth}
						\centering
						\includegraphics[width=.8\linewidth]{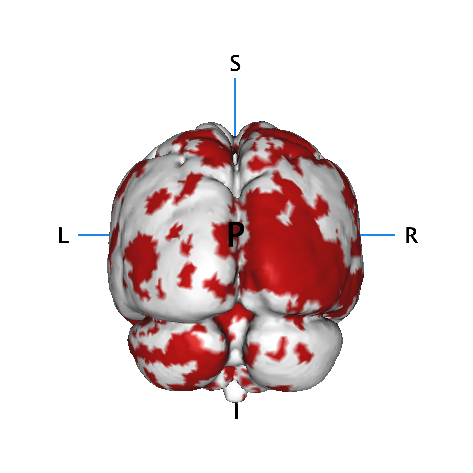}
						\caption{MDR (posterior)}
						\label{fig:sfig10}
					\end{subfigure}
					\begin{subfigure}{.32\textwidth}
						\centering
						\includegraphics[width=.8\linewidth]{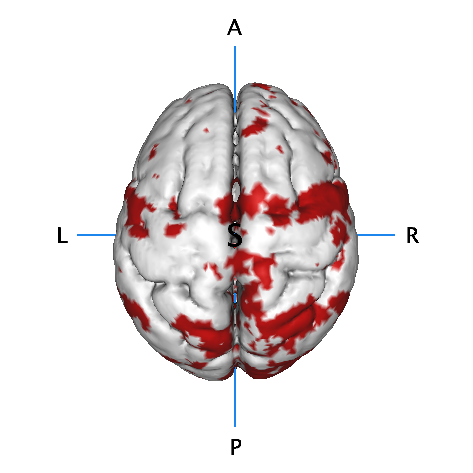}
						\caption{MDR (superior)}
						\label{fig:sfig11}
					\end{subfigure}
					\begin{subfigure}{.32\textwidth}
						\centering
						\includegraphics[width=.8\linewidth]{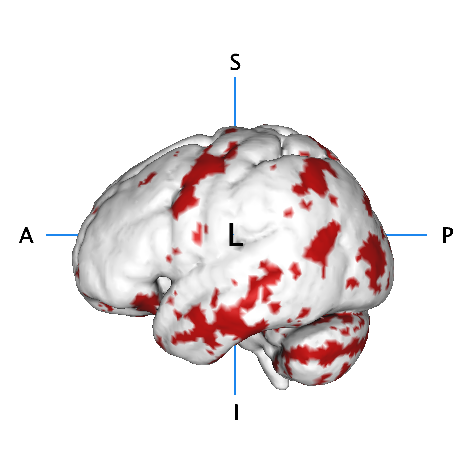}
						\caption{MDR (left)}
						\label{fig:sfig12}
					\end{subfigure}
					\caption{The selected voxels from BH-FDR, DCOE, AFNC, and MDR for a single participant (sub-05).}
					\label{fig:fmri}
				\end{figure}

				\autoref{fig:fmri} illustrates the selected voxels of each method in the image of a single participant (sub-5) from posterior, superior, and left views. The figure is generated using the Multi-image Analysis GUI (http://ric.uthscsa.edu/mango/). 
				We look into several regions that are known to be associated with saccadic eye movements. 
				First, the visual cortex (VC) on occipital lobe in the posterior region of the brain, as indicated in \autoref{fig:fnpc_p}, is the primary cortical region that receives, integrates, and processes visual information \citep{bodis1997functional}. 
				We can see that all four methods have identified voxels in VC. However, the results of DCOE, AFNC and MDR seem to match the VC region much better than that of BH-FDR.    
				From the superior and left views, it shows that BH-FDR has only a few or no discoveries in the frontal eye fields (FEF) located in Brodmann area 8, the supplementary eye fields (SEF) located in Broadmann area 6, and the posterior parietal cortex (PPC). The locations of FEF, SEF, and PPC are indicated in \autoref{fig:fnp_s} and  \autoref{fig:fnp_l}. 
				FEF and SEF are believed to play important roles in visual attention and eye movements as electrical
				stimulation of these areas evokes eye movements \citep{bruce1985primateI, bruce1985primateII}. PPC, on the other hand,  is related to decision making and saccades \citep{goldberg2002role, schluppeck2005topographic}.
				It can be seen that DCOE, AFNC, and MDR have better power for identifying signal voxels in FEF, SEF, and PPC.  
				Among these three methods, DCOE selects the least scattered or isolated voxels that are not functionally relevant to saccades. 
				
\section{Conclusion and Discussion} \label{sec:conclusion}

In this paper, we develop a new analytic framework to perform inference on signals that are detectable yet unidentifiable. A new user-adaptive method is proposed to control FNP at a user-specified level and, at the same time, regulate the unnecessary false positives to achieve the target FNP level. 
The new method is developed under arbitrary covariance dependence calibrated through a dependence parameter whose scale is compatible with the existing phase diagram in high-dimensional sparse inference. Utilizing the new calibration, we are able to explicate the joint effects of covariance dependence, signal sparsity, and signal intensity and interpret the results through a new phase digram to gain important insight.

We demonstrate the finite sample performance of the dual control method in simulation and apply the method to identify functionally relevant regions for saccadic eye movement using fMRI data. Its results are compared with those of other methods and with regions that are known to be associated with saccade. The new method seems to benefit from its dual control property and exhibits a nice balance in identifying signal voxels in the functionally relevant regions and avoiding the scattered noise voxels. Follow up research in applying the new method to presurgical planning to address the major concern of false negative control will be explored in the future. 

The new method will be useful in applications where false negative results substantially hinder scientific investigations. Although such examples are abundant in real life, method development has been quite limited. Our study provides an insightful approach to bridge the important methodological gap. Moreover, as the new method provides a  more lenient cutoff compared to the stringent cutoffs of multiple testing procedures, candidate variables can be divided into three groups with the two cutoffs. This allows practitioners to design different follow-up studies for candidates in each group. Developing innovative strategies taking advantages of the three-group classification is of great interest in future research.

				\section*{Acknowledgment}
				The authors are grateful to Dr. Bertrand Thirion and Dr. Ana Lu{\'i}sa Grilo Pinho for providing background information of the IBC dataset. 
			
				\section{Appendix}
				
				\begin{small}

				This section provides the proofs of Theorem \ref{thm:FNPcontrol} and Proposition \ref{prop:estimated_s}.  We will frequently use the following result on Mill's ratio:
				\[
				\bar \Phi(x) \le x^{-1} \phi(x) \qquad \mbox{for any} \quad x>0.
				\] 
				The symbol $C$ denotes a genetic, finite constant whose value can be different at different occurrences.

				\subsection{Proof of \autoref{thm:FNPcontrol}}

First, we show that $\widehat{\mbox{FNP}}(t)$ can accurately approximate $\mbox{FNP}(t)$ when $t$ is large enough. Proof of the following lemma is provided in Section \ref{sec:proof_consistency}.

\begin{lemma} \label{lemma:consistency}
Consider model (\ref{def:multiNorm}). For any $t=t_p \ge \mu_{min}$, where $\mu_{min}$ is defined in (\ref{def:t_min}), we have
	\begin{equation} \label{eq:consistency}
		|\widehat{\mbox{FNP}}(t) - \mbox{FNP}(t)|=o_P(1).
	\end{equation}
\end{lemma}

Then, by Markov's inequality, for a fixed constant $a>0$,
				\begin{eqnarray*}
					P(\mbox{FNP}(\mu_{min})>a) & = & P( s^{-1}\sum_{j\in I_1}1_{\{Z_j \leq \mu_{min}\}} > a) \le \frac{1}{as}\sum_{j\in I_1} P(Z_j \le \mu_{min}) \\ & \leq & \frac{1}{a} \max_{j\in I_1} P(N(0,1) \le \mu_{min}- A_j) \le \frac{1}{a} \max_{j\in I_1} P(N(0,1) \le \mu_{min}- A_{min}) =o(1),
				\end{eqnarray*}
				where the last step is by the condition on $A_{\min}$. Therefore, $\mbox{FNP}(\mu_{min})=o_P(1)$. By \autoref{lemma:consistency}, we have  $|\widehat{\mbox{FNP}}(\mu_{min})-\mbox{FNP}(\mu_{min})|=o_P(1)$, which implies  $\widehat{\mbox{FNP}}(\mu_{min})=o_P(1)$. Then,  by the construction of $\hat{t}(\beta)$, 
				we have $P(\hat{t} (\beta) \geq \mu_{min})\rightarrow 1$. Consequently, since $\hat{t} (\beta)$ is random, 
				\begin{eqnarray*}
					& & \lim_{p\to \infty} P(|\widehat{\mbox{FNP}}(\hat{t}(\beta))- \mbox{FNP}(\hat{t}(\beta))| >a) \\ 
					& = & \lim_{p\to \infty} E\left[ E\left( 1\{|\widehat{\mbox{FNP}}(\hat{t}(\beta))- \mbox{FNP}(\hat{t}(\beta))| >a\} | \hat{t}(\beta)\right)\right] \\
					& = & E\left[ \lim_{p\to \infty}  E\left( 1\{|\widehat{\mbox{FNP}}(\hat{t}(\beta))- \mbox{FNP}(\hat{t}(\beta))| >a\} | \hat{t}(\beta)\right)\right] = 0,
				\end{eqnarray*}
				where the second step above is by dominated convergence theorem and the last step is by \autoref{lemma:consistency}. Therefore, $|\widehat{\mbox{FNP}}(\hat{t}(\beta))- \mbox{FNP}(\hat{t}(\beta))|=o_P(1)$. Now, since $\widehat{\mbox{FNP}}(\hat{t}(\beta))<\beta$ almost surely by the construction of $\hat{t}(\beta)$, claim in (\ref{eq:FNPcontrol_1}) follows. 
				
				Next, let's consider the claim in (\ref{eq:FNPcontrol_2}). Because $\tilde{t} > \hat{t}(\beta)$, the construction of  $\hat{t}(\beta)$ implies that $\widehat{\mbox{FNP}}(\tilde{t}) \ge \beta$ almost surely. On the other hand, because 
				$\tilde{t} > \hat{t}(\beta) \ge \mu_{min}$ with probability tending to 1, similar arguments as the above lead to   $|\widehat{\mbox{FNP}}(\tilde{t})- \mbox{FNP}(\tilde{t})|=o_P(1)$. Therefore, the claim in (\ref{eq:FNPcontrol_2}) follows. 
				\\
				
\subsection{Proof of \autoref{lemma:consistency}} \label{sec:proof_consistency}

For notation simplicity, let 
\begin{equation} \label{def:B(t)}
	B(t) = 1-\mbox{R}(t)/s+ (p-s) \Bar{\Phi}(t)/s. 
\end{equation}
Then $\widehat{\mbox{FNP}}(t) = \max\{B(t), ~0\}$, and it is sufficient to show that
\begin{equation} \label{01}
	|B(t) - \mbox{FNP}(t)| = o_ p(1) \qquad \mbox{when} \quad B(t) \ge 0
\end{equation}
and 
\begin{equation} \label{02}
	\mbox{FNP}(t) = o_ p(1)  \qquad \mbox{when} \quad B(t) < 0. 
\end{equation}

Consider (\ref{01}) first. By the definitions of $B(t)$ and $\mbox{FNP}(t)$,  
\[
|B(t) - \mbox{FNP}(t)| = |s^{-1}(\mbox{R}(t)-(p-s)\Bar{\Phi}(t)) - s^{-1}(\mbox{R}(t)-\mbox{FP}(t))| = s^{-1}|\mbox{FP}(t) - (p-s)\Bar{\Phi}(t)|. 
\]
Therefore, it is sufficient to show
\begin{equation} \label{0}
	s^{-1}|\mbox{FP}(t) - (p-s)\Bar{\Phi}(t)| = o_P(1).  
\end{equation}
Recall the definition of $\mu_1$ and $\mu_2$ in (\ref{def:t_min}). The following proof is composed of two parts. The first part assumes $t \ge \mu_1$ and the second part assumes $\mu_2 \le t < \mu_1$.

Consider the first part. It's sufficient to show $s^{-1}\mbox{FP}(t)=o_P(1)$ and $s^{-1} (p-s) \Bar{\Phi}(t)=o(1)$ with $t \ge \mu_1$. By Mill's ratio and the re-parameterization of $s$ in $\gamma$,
\begin{equation*}
	s^{-1}(p-s)\Bar{\Phi}(t)\leq \frac{Cpe^{-t^2/2}}{ts} \le  \frac{C} {\sqrt{ \log{p}}}=o(1).
\end{equation*}
On the other hand, for a fixed constant $a>0$,
\begin{equation*}
	P(s^{-1}\mbox{FP}(t)>a) \leq \frac{E(\mbox{FP}(t))}{as} \leq \frac{p \max_{j\in I_0}P( Z_j >t)}{as} \le   \frac{Cp\Bar{\Phi}(t)}{s} = o(1).
\end{equation*}
Therefore, the claim in (\ref{0}) is justified for $t \ge \mu_1 = \sqrt{2\gamma \log{p}}$.

Next we present the second part of the proof with $\mu_2 \le t < \mu_1$. Define
\begin{equation*}
	D_p=s^{-2}e^{-t^2/2}\lVert\mathbf{\Sigma}\rVert_1\log{p}. 
\end{equation*}
By the condition $t \ge \mu_2 $ and the re-parameterizations of $\lVert\mathbf{\Sigma}\rVert_1$ in $\eta$ and $s$ in $\gamma$,  it can be shown that 
\begin{eqnarray*}
	D_p  = p^{2\gamma-\eta}e^{-t^2/2} \log{p} \le \log ^{-1} p = o(1). 
\end{eqnarray*}
Then, (\ref{0}) is implied by
\begin{equation} \label{0.1}
	s^{-1}|\mbox{FP}({t})-(p-s)\Bar{\Phi}(t)| = o_P(\sqrt{D_p}). 
\end{equation}
Apply Chebyshev's inequality,
\begin{equation} \label{0.2}
	P(s^{-1}|\mbox{FP}(t)-(p-s)\Bar{\Phi}(t)|>\sqrt{D_p})
	\leq\frac{ Var(\mbox{FP}(t))}{s^2D_p}.
\end{equation}
We have the following lemma for the order of $Var(\mbox{FP}(t))$ under dependence. The proof is essentially the same as that of Lemma 5.2 in 
\cite{jeng2021estimating}. We omit the details to save space. 

\begin{lemma}\label{lemma:Var(FP)} 
	Consider  model (\ref{def:multiNorm}). Denote $\mathbf{\Sigma}_0$ as the correlation matrix of $Z_j, j \in I_0$. We have 
	\begin{equation*}
		Var(\sum_{j \in I_0} 1_{\{Z_j>t\}})=O(e^{-t^2/2}\lVert\mathbf{\Sigma_0}\rVert_1).
	\end{equation*}
\end{lemma}

Therefore, $Var(\mbox{FP}(t))=Var(\sum_{j\in I_0}1_{\{Z_j >t\}}) \le C e^{-t^2/2}\lVert\mathbf{\Sigma}_0\rVert_1 \le  C e^{-t^2/2}\lVert\mathbf{\Sigma}\rVert_1$, and it follows that
\begin{equation} \label{0.3}
	\frac{Var(\mbox{FP}(t))}{s^2D_p}=o(1)
\end{equation}
by the definition of $D_p$. Combining (\ref{0.2}) and (\ref{0.3}) gives (\ref{0.1}), which justifies the claim in (\ref{0}) for $\mu_2 \le t < \mu_1$.

Finally, consider (\ref{02}). It can be shown that $B(t) <0$ implies  $1-\mbox{FP}(t)/s - \mbox{TP}(t) / s + (p-s) \Bar{\Phi}(t)/s <0$, which, combined with FNP$(t) = 1- \mbox{TP}(t)/s$, further implies 
\[
\mbox{FNP}(t) < s^{-1}(\mbox{FP}(t) - (p-s)\Bar{\Phi}(t)).  
\]
Since the order of the right hand side has been derived in (\ref{0}), (\ref{02}) follows.
\\

				\subsection{Proof of Proposition \ref{prop:estimated_s}}
				
				The consistency of $\hat \pi_{}^*$ is based on the following two lemmas, which are implied by Theorem 2.2 and 2.3 in \cite{jeng2021estimating}. 
				
				\begin{lemma} \label{lemma:discrete05}
					Consider model (\ref{def:multiNorm}). Let $\delta(t) = [\bar \Phi(t)]^{1/2}$. Then, there exists a bounding sequence 
					$c_{p, \delta}^* = O\left(\sqrt{\bar \rho \cdot  \log p}\right)$,
					and the corresponding estimator $\hat{\pi}^*_{\delta}$  satisfies $ P(\hat \pi^*_{\delta} < \pi) \to 1$. Moreover, for $\pi$ satisfying $0 < \pi \ll  1$,	
					if $A_{min} \gg 1$ and 	
					\begin{equation} \label{cond:theta_05}
						A_{min} - \bar \Phi^{-1}\left( {\pi^2 \over \bar \rho \cdot \log p } \right) \to \infty,
					\end{equation}
					then $P(\hat \pi^*_{\delta} > (1-\epsilon) \pi) \to 1$ for any constant $\epsilon>0$.
				\end{lemma}

				\begin{lemma} \label{lemma:discrete1}
					Consider model (\ref{def:multiNorm}).
					Let $\delta(t) = \bar \Phi(t)$. Then, there exists a bounding sequence $c_{p, \delta}^* = O(\sqrt{\log p})$, and the corresponding estimator $\hat{\pi}^*_{\delta}$ satisfies $ P(\hat \pi^*_{\delta} < \pi) \to 1$. Moreover, for $\pi$ satisfying $0 < \pi \ll  1$,  if $A_{min} \gg 1$ and 
					\begin{equation} \label{cond:theta_1}
						A_{min} - \bar \Phi^{-1}\left({\pi \over \sqrt{\log p} }\right) \to \infty.
					\end{equation}
					then $P(\hat \pi^*_{\delta} > (1-\epsilon) \pi) \to 1$ for any constant $\epsilon>0$.	
				\end{lemma}

				Recall that $\pi = p^{-\gamma}$, $\gamma \in (0, 1)$, and $\bar \rho = p^{-\eta}, \eta \in [0,1]$. Recall $\mu_1$ and $\mu_2$ in (\ref{def:t_min}). Direct calculation shows that when $A_{min} - \mu_2 \to \infty$, the condition in (\ref{cond:theta_05}) is satisfied, and when  $A_{min} - \mu_1 \to \infty$, the condition in (\ref{cond:theta_1}) is satisfied. Therefore, given $A_{min} - \mu_{min} \to \infty$, the consistency of the estimator  $\hat \pi_{}^* = \max\{\hat \pi_{0.5}^*, \hat \pi_{1}^*\}$  holds as in (\ref{eq:hat_pi}).

				Next consider the claims in (\ref{eq:FNPcontrol_1s}) and (\ref{eq:FNPcontrol_2s}). Similar arguments as in the proof of \autoref{thm:FNPcontrol} can be applied, and we only need to show that \autoref{lemma:consistency} continues to hold with $\widehat{\mbox{FNP}}(t)$ replaced by $\widehat{\mbox{FNP}}_{\hat s}(t)$. Given the result in (\ref{eq:consistency}), it is sufficient to show
				$|\widehat{\mbox{FNP}}_{\hat s}(t) - \widehat{\mbox{FNP}}(t) | = o_P(1)$ for $t=t_p \ge \mu_{min}$, which, by the definition of $B(t)$ in (\ref{def:B(t)}),  is implied by 
				\begin{equation} \label{3}
					|B_{\hat s}(t) - B(t) | = o_P(1)  \quad \mbox{for} \quad t=t_p \ge \mu_{min}.
				\end{equation}  
				By direct calculation, 
				\begin{eqnarray} \label{3.0}
					|B_{\hat s}(t) - B(t) | & = & |(\hat s^{-1} - s^{-1}) (\mbox{R}(t) - p \bar \Phi(t))| \nonumber \\
					& \le & |\hat s^{-1} - s^{-1}| \cdot \mbox{TP}(t) + |\hat s^{-1} - s^{-1}| \cdot |\mbox{FP}(t) - p \bar \Phi(t)|. 
				\end{eqnarray}
				Given $P((1-\delta) < \hat s/s  < 1)\rightarrow 1$ for any $\delta>0$, it can be shown that 
				\[
				P(|\hat s^{-1} - s^{-1}| < \frac{\delta}{1-\delta} s^{-1}) \to 1. 
				\]
				On the other hand, $TP(t) \le s$ almost surely. Then it follows that the first term in (\ref{3.0}),
				\[
				|\hat s^{-1} - s^{-1}| \cdot \mbox{TP}(t) = o_p(1).
				\]
				For the second term in (\ref{3.0}), 
				\[
				|\hat s^{-1} - s^{-1}| \cdot |\mbox{FP}(t) - p \bar \Phi(t)| < \frac{\delta}{1-\delta} s^{-1} |\mbox{FP}(t) - p \bar \Phi(t)|) \le \frac{\delta}{1-\delta} \left(s^{-1} |\mbox{FP}(t) - (p-s) \bar \Phi(t)|) + \bar \Phi(t)\right)
				\]
				with probability tending 1, where $\bar \Phi(t) = o(1)$ for $t \ge \mu_{min}$, and it has been shown as for (\ref{0}) that 
				$s^{-1}|\mbox{FP}(t) - (p-s)\Bar{\Phi}(t)| = o_P(1)$ for $t \ge \mu_{min}$. Then it follows that
				\[
				|\hat s^{-1} - s^{-1}| \cdot |\mbox{FP}(t) - p \bar \Phi(t)| = o_P(1).
				\] 
				Summing up the above gives (\ref{3}). 
				\\

			\end{small}
				
				


\bibliographystyle{chicago}
\bibliography{FNP_reference}

\begin{thebibliography}{}

\bibitem[\protect\citeauthoryear{Arias-Castro, Cand{\`e}s, and
  Plan}{Arias-Castro et~al.}{2011}]{arias2011global}
Arias-Castro, E., E.~J. Cand{\`e}s, and Y.~Plan (2011).
\newblock Global testing under sparse alternatives: Anova, multiple comparisons
  and the higher criticism.
\newblock {\em The Annals of Statistics\/}~{\em 39\/}(5), 2533--2556.

\bibitem[\protect\citeauthoryear{{Benjamini} and {Hochberg}}{{Benjamini} and
  {Hochberg}}{1995}]{benjamini1995}
{Benjamini}, Y. and Y.~{Hochberg} (1995).
\newblock Controlling the false discovery rate: a practical and powerful
  approach to multiple testing.
\newblock {\em J. R. Statist. Soc. Ser. B\/}~{\em 57\/}(1), 289--300.

\bibitem[\protect\citeauthoryear{Bodis-Wollner, Bucher, Seelos, Paulus, Reiser,
  and Oertel}{Bodis-Wollner et~al.}{1997}]{bodis1997functional}
Bodis-Wollner, I., S.~Bucher, K.~Seelos, W.~Paulus, M.~Reiser, and W.~Oertel
  (1997).
\newblock Functional mri mapping of occipital and frontal cortical activity
  during voluntary and imagined saccades.
\newblock {\em Neurology\/}~{\em 49\/}(2), 416--420.

\bibitem[\protect\citeauthoryear{Bruce and Goldberg}{Bruce and
  Goldberg}{1985}]{bruce1985primateI}
Bruce, C.~J. and M.~E. Goldberg (1985).
\newblock Primate frontal eye fields. i. single neurons discharging before
  saccades.
\newblock {\em Journal of neurophysiology\/}~{\em 53\/}(3), 603--635.

\bibitem[\protect\citeauthoryear{Bruce, Goldberg, Bushnell, and Stanton}{Bruce
  et~al.}{1985}]{bruce1985primateII}
Bruce, C.~J., M.~E. Goldberg, M.~C. Bushnell, and G.~B. Stanton (1985).
\newblock Primate frontal eye fields. ii. physiological and anatomical
  correlates of electrically evoked eye movements.
\newblock {\em Journal of neurophysiology\/}~{\em 54\/}(3), 714--734.

\bibitem[\protect\citeauthoryear{Cai, Jin, and Low}{Cai
  et~al.}{2007}]{cai2007estimation}
Cai, T., J.~Jin, and M.~Low (2007).
\newblock Estimation and confidence sets for sparse normal mixtures.
\newblock {\em The Annals of Statistics\/}~{\em 35\/}(6), 2421--2449.

\bibitem[\protect\citeauthoryear{Cai, Jeng, and Jin}{Cai
  et~al.}{2011}]{tony2011optimal}
Cai, T.~T., X.~J. Jeng, and J.~Jin (2011).
\newblock Optimal detection of heterogeneous and heteroscedastic mixtures.
\newblock {\em Journal of the Royal Statistical Society: Series B (Statistical
  Methodology)\/}~{\em 73\/}(5), 629--662.

\bibitem[\protect\citeauthoryear{Cai and Jin}{Cai and
  Jin}{2010}]{cai2010optimal}
Cai, T.~T. and J.~Jin (2010).
\newblock Optimal rates of convergence for estimating the null density and
  proportion of nonnull effects in large-scale multiple testing.
\newblock {\em The Annals of Statistics\/}, 100--145.

\bibitem[\protect\citeauthoryear{Cai and Sun}{Cai and
  Sun}{2017a}]{cai2017large}
Cai, T.~T. and W.~Sun (2017a).
\newblock Large-scale global and simultaneous inference: Estimation and testing
  in very high dimensions.
\newblock {\em Annual Review of Economics\/}~{\em 9}, 411--439.

\bibitem[\protect\citeauthoryear{Cai and Sun}{Cai and
  Sun}{2017b}]{cai2016optimal}
Cai, T.~T. and W.~Sun (2017b).
\newblock Optimal screening and discovery of sparse signals with applications
  to multistage high-throughput studies.
\newblock {\em Journal of the Royal Statistical Society: Series B\/}~{\em
  79\/}(1), 197--223.

\bibitem[\protect\citeauthoryear{Chen, Li, and Zhong}{Chen
  et~al.}{2019}]{chen2019two}
Chen, S.~X., J.~Li, and P.-S. Zhong (2019).
\newblock Two-sample and anova tests for high dimensional means.
\newblock {\em The Annals of Statistics\/}~{\em 47\/}(3), 1443--1474.

\bibitem[\protect\citeauthoryear{Donoho and Jin}{Donoho and
  Jin}{2004}]{donoho2004higher}
Donoho, D. and J.~Jin (2004).
\newblock Higher criticism for detecting sparse heterogeneous mixtures.
\newblock {\em The Annals of Statistics\/}~{\em 32\/}(3), 962--994.

\bibitem[\protect\citeauthoryear{Donoho and Jin}{Donoho and
  Jin}{2015}]{donoho2015special}
Donoho, D. and J.~Jin (2015).
\newblock Special invited paper: Higher criticism for large-scale inference,
  especially for rare and weak effects.
\newblock {\em Statistical Science\/}, 1--25.

\bibitem[\protect\citeauthoryear{Fowlkes and Mallows}{Fowlkes and
  Mallows}{1983}]{fowlkes1983method}
Fowlkes, E.~B. and C.~L. Mallows (1983).
\newblock A method for comparing two hierarchical clusterings.
\newblock {\em Journal of the American statistical association\/}~{\em
  78\/}(383), 553--569.

\bibitem[\protect\citeauthoryear{Genovese and Wasserman}{Genovese and
  Wasserman}{2002}]{genovese2002operating}
Genovese, C. and L.~Wasserman (2002).
\newblock Operating characteristics and extensions of the false discovery rate
  procedure.
\newblock {\em Journal of the Royal Statistical Society: Series B (Statistical
  Methodology)\/}~{\em 64\/}(3), 499--517.

\bibitem[\protect\citeauthoryear{Genovese and Wasserman}{Genovese and
  Wasserman}{2004}]{GW04}
Genovese, C. and L.~Wasserman (2004).
\newblock A stochastic process approach to false discovery control.
\newblock {\em Annals of Statistics\/}~{\em 32}, 1035--1061.

\bibitem[\protect\citeauthoryear{Goldberg, Bisley, Powell, Gottlieb, and
  Kusunoki}{Goldberg et~al.}{2002}]{goldberg2002role}
Goldberg, M.~E., J.~Bisley, K.~D. Powell, J.~Gottlieb, and M.~Kusunoki (2002).
\newblock The role of the lateral intraparietal area of the monkey in the
  generation of saccades and visuospatial attention.
\newblock {\em Annals of the New York Academy of Sciences\/}~{\em 956\/}(1),
  205--215.

\bibitem[\protect\citeauthoryear{Halkidi, Batistakis, and Vazirgiannis}{Halkidi
  et~al.}{2001}]{halkidi2001clustering}
Halkidi, M., Y.~Batistakis, and M.~Vazirgiannis (2001).
\newblock On clustering validation techniques.
\newblock {\em Journal of intelligent information systems\/}~{\em 17\/}(2-3),
  107--145.

\bibitem[\protect\citeauthoryear{Ingster}{Ingster}{1994}]{ingster1994minimax}
Ingster, Y.~I. (1994).
\newblock Minimax detection of a signal in ? p metrics.
\newblock {\em Journal of Mathematical Sciences\/}~{\em 68\/}(4), 503--515.

\bibitem[\protect\citeauthoryear{Jeng}{Jeng}{2021}]{jeng2021estimating}
Jeng, X.~J. (2021).
\newblock Estimating the proportion of signal variables under arbitrary
  covariance dependence.
\newblock {\em arXiv preprint arXiv:2102.09053\/}.

\bibitem[\protect\citeauthoryear{Jeng and Chen}{Jeng and
  Chen}{2019}]{JengChen2019}
Jeng, X.~J. and X.~Chen (2019).
\newblock Variable selection via adaptive false negative control in linear
  regression.
\newblock {\em Electron. J. Statist.\/}~{\em 13\/}(2), 5306--5333.

\bibitem[\protect\citeauthoryear{Jeng, Daye, Lu, and Tzeng}{Jeng
  et~al.}{2016}]{jeng2016rare}
Jeng, X.~J., Z.~J. Daye, W.~Lu, and J.-Y. Tzeng (2016).
\newblock Rare variants association analysis in large-scale sequencing studies
  at the single locus level.
\newblock {\em PLoS computational biology\/}~{\em 12\/}(6).

\bibitem[\protect\citeauthoryear{Jeng, Zhang, and Tzeng}{Jeng
  et~al.}{2019}]{jeng2019efficient}
Jeng, X.~J., T.~Zhang, and J.-Y. Tzeng (2019).
\newblock Efficient signal inclusion with genomic applications.
\newblock {\em Journal of the American Statistical Association\/}~{\em 114},
  1787--1799.

\bibitem[\protect\citeauthoryear{Ji and Jin}{Ji and Jin}{2012}]{ji2012ups}
Ji, P. and J.~Jin (2012).
\newblock Ups delivers optimal phase diagram in high-dimensional variable
  selection.
\newblock {\em The Annals of Statistics\/}~{\em 40\/}(1), 73--103.

\bibitem[\protect\citeauthoryear{Ji and Zhao}{Ji and Zhao}{2014}]{ji2014rate}
Ji, P. and Z.~Zhao (2014).
\newblock Rate optimal multiple testing procedure in high-dimensional
  regression.
\newblock {\em arXiv preprint arXiv:1404.2961\/}.

\bibitem[\protect\citeauthoryear{Jin, Ke, and Wang}{Jin
  et~al.}{2017}]{jin2017phase}
Jin, J., Z.~T. Ke, and W.~Wang (2017).
\newblock Phase transitions for high dimensional clustering and related
  problems.
\newblock {\em The Annals of Statistics\/}~{\em 45\/}(5), 2151--2189.

\bibitem[\protect\citeauthoryear{Meinshausen and Rice}{Meinshausen and
  Rice}{2006}]{MR06}
Meinshausen, N. and J.~Rice (2006).
\newblock Estimating the proportion of false null hypotheses among a large
  number of independently tested hypotheses.
\newblock {\em Ann. Statist.\/}~{\em 34\/}(1), 373--393.

\bibitem[\protect\citeauthoryear{Pinho, Amadon, Ruest, Fabre, Dohmatob,
  Denghien, Ginisty, Becuwe-Desmidt, Roger, Laurier, et~al.}{Pinho
  et~al.}{2018}]{pinho2018individual}
Pinho, A.~L., A.~Amadon, T.~Ruest, M.~Fabre, E.~Dohmatob, I.~Denghien,
  C.~Ginisty, S.~Becuwe-Desmidt, S.~Roger, L.~Laurier, et~al. (2018).
\newblock Individual brain charting, a high-resolution fmri dataset for
  cognitive mapping.
\newblock {\em Scientific data\/}~{\em 5}, 180105.

\bibitem[\protect\citeauthoryear{Sarkar}{Sarkar}{2006}]{sarkar2006false}
Sarkar, S.~K. (2006).
\newblock False discovery and false nondiscovery rates in single-step multiple
  testing procedures.
\newblock {\em The Annals of Statistics\/}~{\em 34\/}(1), 394--415.

\bibitem[\protect\citeauthoryear{Schluppeck, Glimcher, and Heeger}{Schluppeck
  et~al.}{2005}]{schluppeck2005topographic}
Schluppeck, D., P.~Glimcher, and D.~J. Heeger (2005).
\newblock Topographic organization for delayed saccades in human posterior
  parietal cortex.
\newblock {\em Journal of neurophysiology\/}~{\em 94\/}(2), 1372--1384.

\bibitem[\protect\citeauthoryear{Visscher, Wray, Zhang, Sklar, McCarthy, Brown,
  and Yang}{Visscher et~al.}{2017}]{visscher201710}
Visscher, P.~M., N.~R. Wray, Q.~Zhang, P.~Sklar, M.~I. McCarthy, M.~A. Brown,
  and J.~Yang (2017).
\newblock 10 years of gwas discovery: biology, function, and translation.
\newblock {\em The American Journal of Human Genetics\/}~{\em 101\/}(1), 5--22.
  PMCID: PMC5501872.

\bibitem[\protect\citeauthoryear{Westfall and Young}{Westfall and
  Young}{1993}]{westfall1993resampling}
Westfall, P.~H. and S.~S. Young (1993).
\newblock {\em Resampling-based multiple testing: Examples and methods for
  p-value adjustment}, Volume 279.
\newblock John Wiley \& Sons.

\end{thebibliography}

 
\end{document}